\documentclass[aps,prb,twocolumn,showpacs,superscriptaddress]{revtex4-2}

\usepackage{graphicx}
\usepackage{amsmath}
\usepackage{multirow}
\usepackage{tabularx}
\usepackage{color}
\usepackage{float}
\usepackage{appendix}
\usepackage[colorlinks,linkcolor=blue,anchorcolor=blue,citecolor=blue]{hyperref}

\begin{document}
\title{Phase diagram and critical behavior of Hubbard model on the square-hexagon-octagon lattice}

\author{Xinwei Jia}
\affiliation{\mbox{Guangdong Provincial Key Laboratory of Magnetoelectric Physics and Devices,}
	\mbox{State Key Laboratory of Optoelectronic Materials and Technologies,}
	\mbox{Center for Neutron Science and Technology,}
	\mbox{School of Physics, Sun Yat-sen University, Guangzhou, 510275, China}}

\author{Dao-Xin Yao}
\email{yaodaox@mail.sysu.edu.cn}
\affiliation{\mbox{Guangdong Provincial Key Laboratory of Magnetoelectric Physics and Devices,}
	\mbox{State Key Laboratory of Optoelectronic Materials and Technologies,}
	\mbox{Center for Neutron Science and Technology,}
	\mbox{School of Physics, Sun Yat-sen University, Guangzhou, 510275, China}}
\affiliation{\mbox{International Quantum Academy, Shenzhen 518048, China}}

\author{Han-Qing Wu}
\email{wuhanq3@mail.sysu.edu.cn}
\affiliation{\mbox{Guangdong Provincial Key Laboratory of Magnetoelectric Physics and Devices,}
	\mbox{State Key Laboratory of Optoelectronic Materials and Technologies,}
	\mbox{Center for Neutron Science and Technology,}
	\mbox{School of Physics, Sun Yat-sen University, Guangzhou, 510275, China}}

\begin{abstract}

Employing the projective formalism of determinant quantum Monte Carlo (DQMC) simulations, we meticulously explore the ground-state phase diagram and critical behavior of the half-filled Hubbard model on a square-hexagon-octagon (SHO) lattice. 
This lattice, a two-dimensional (2D) structure comprising squares, hexagons, and octagons, is representative of the biphenylene network (BPN). 
Our findings reveal an intriguing ground-state phase diagram, featuring an antiferromagnetic (AFM) Mott insulating phase enveloped by three valence-bond-solid-like (VBS-like) insulating phases. 
Analyzing the single-particle gap, spin gap, and single-particle spectral function, we observe that the metallic state in the noninteracting case becomes unstable under the influence of Hubbard $U$. 
This interaction drives the system into a hexagon insulating phase before transitioning into an AFM Mott insulating phase. 
To quantify the critical exponents, we use finite-size scaling techniques. 
The critical exponents of quantum critical points between the AFM Mott insulating phase and two insulating phases, plaquette insulator, and ethylene insulator, closely align with the 3D O(3) universality class. 
However, the critical exponents of quantum critical points between the hexagon insulating phase and the AFM Mott insulating phase deviate from the 3D O(3) universality class. 
This deviation can be attributed to the coupling effect between the fluctuations of magnetic order parameters and very low-energy fermionic excitations. 
Our comprehensive study not only advances the understanding of correlation effects on the SHO lattice but also sheds light on the less-explored critical exponents in a weakly insulating quantum critical point.
      
\end{abstract}

\maketitle

\section{INTRODUCTION}\label{INTRODUCTION}

The discovery of single-layer graphene \cite{doi:10.1126/science.1102896} and its novel quantum physical phenomena \cite{novoselov2005two, zhang2005experimental}, like the massless Dirac fermion, have attracted wide attention to the research of two-dimensional (2D) materials. Besides graphene, single or multi-layer 2D structures have also been discovered in other materials, such as $MoS{_2}$ \cite{radisavljevic2011single}, they all share the hexagon lattice structure which contributes to the unique physical properties. 
In order to enrich the family of low-dimensional graphenic structures, many efforts have been made by researchers and embedding non-hexagonal rings into $sp^2$-hybridized carbon networks is considered as a promising strategy. 
In Ref.\cite{10.1038/ncomms14924}, they proposed an on-surface synthesis of graphene-like nanoribbons with periodically embedded four- and eight-membered rings, making it possible to unveil the properties induced by non-hexagonal rings. Subsequently, Ref.\cite{10.1126/science.abg4509} report the growth of an ultraflat biphenylene network with periodically arranged four-, six-, and eight-membered rings of $sp^2$-hybridized carbon atoms. Density functional theory (DFT) calculations of the biphenylene sheet \cite{doi:10.1021/nn100758h, GE201697, Ye_2023} show that it has peculiar electronic properties and the potential to achieve high-temperature superconductivity \cite{Ye_2023, JunLi17403, PhysRevB.102.174509, PhysRevB.99.184506}. In addition, the topological magnons and the Einstein–de Haas effect were studied on the square-hexagon-octagon lattice \cite{PhysRevB.108.144407}.

The impact of strong correlation effects of Hubbard model on hexagonal honeycomb lattice has been deeply studied, revealing very interesting phenomena such as metal-insulator transition of Dirac fermions, antiferromagnetism and superconductivity \cite{PhysRevLett.97.230404, PhysRevB.105.214510, CICHY2024171522, meng2010quantum, sorella2012absence, Otsuka_2022}.
To the best of our knowledge, no large-scale numerical simulation has been performed for the ground-state phase diagram and critical behavior of Hubbard model on square-hexagon-octagon (SHO) lattice.
Motivated by the biphenylene network, we want to study the correlation effects of Hubbard model on the SHO lattice or its topologically equivalent bond-depleted square (BDS) lattice.
The BDS lattice may be synthetized using transition-metal oxide or simulated in cold atoms more easily.
To investigate the ground-state properties of Hubbard model on the SHO lattice, we employ the determinant quantum Monte Carlo (DQMC) simulations\cite{PhysRevB.71.155115} which is absence of sign problem in the half-filled case.
It uses randomly sampling the auxiliary field based on Markov Chain Monte Carlo (MCMC) algorithm. Ground-state properties can be efficiently obtained through projective formalism \cite{Assaad2008}. 
DQMC has been used to study the Hubbard model on single-layer honeycomb lattice \cite{meng2010quantum, sorella2012absence, Otsuka_2022} and $\pi$-flux model on square lattice \cite{PhysRevX.6.011029, Otsuka_2022}, where the Dirac cone is stable against the local interactions and can only be destroyed by sufficient strong interaction, leading to an AFM Mott insulating phase transition that belongs to the chiral Heisenberg Gross-Neveu universality class \cite{S.Sorella_1992, PhysRevLett.97.146401, PhysRevB.80.075432, PhysRevX.3.031010, PhysRevX.6.011029, arxiv231010601, Otsuka_2022}.

In this paper, we use the projective formalism of DQMC to study the ground-state phase diagram and critical behavior of the Hubbard model on SHO lattice. In the noninteracting tight-binding limit, there are some flat dispersions along high-symmetry lines and some accidental crossing point at the Fermi level. The noninteracting metallic state is unstable to the Hubbard-$U$ interaction, leading to a hexagon VBS-like insulating phase before going into antiferromagnetic Mott phase. 
We also do the finite-size scaling to determine the exact quantum critical points and study critical exponents of quantum phase transitions. 
In addition, we also study the evolution of single-particle spectrum with Hubbard $U$ interaction, which directly displays the correlation effects and can be compared with angle resolved photoemission spectroscopy (ARPES) results. 
Our systematic numerical results are very helpful to understand the correlation effects of the SHO-lattice Hubbard model in transition metal oxides or cold atom systems.

 \begin{figure}[t]
	\centering
	\includegraphics[scale=0.44]{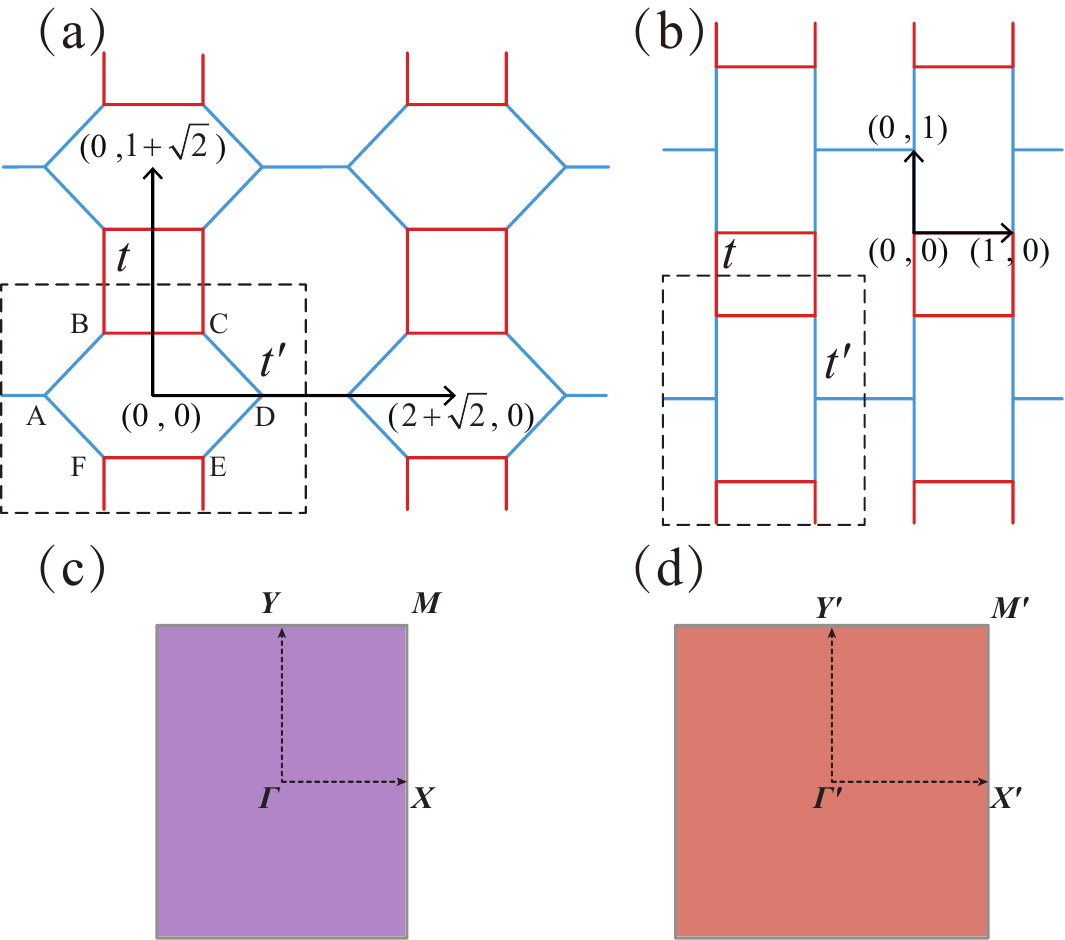}
	\caption{(a) Illustration of the SHO lattice which is composed of squares, hexagons, and octagons. The black dashed box shows one unit cell with six sites. The bonds with red and light blue colors represent two kinds of nearest-neighbor hopping amplitudes $t$ and $t'$, respectively, used in the calculations. (b) The bond-depleted square (BDS) lattice which is topologically equivalent to the SHO lattice. We use this geometry to do the Fourier transform of static spin-spin correlation function with the square-lattice Brillouin zone. (c) The purple region is the first Brillouin zone of the SHO lattice, and high-symmetry points $\varGamma=(0,0), X=(\frac{\pi}{2+\sqrt{2}},0), Y=(0,\frac{\pi}{1+\sqrt{2}}), M=(\frac{\pi}{2+\sqrt{2}},\frac{\pi}{1+\sqrt{2}})$. (d) The red region is the first Brillouin zone of the square lattice, the high-symmetry points $\varGamma^\prime=(0,0), X^\prime=(\pi,0), Y^\prime=(0,\pi), M^\prime=(\pi,\pi)$ are also shown in the figure.}
	\label{fig1-Lattice}
\end{figure}

The rest of this paper is organized as follows. In Sec. ~\ref{MODEL AND METHODS}, we introduce the model and computational methods used in this work. In Sec. ~\ref{RESULT}, we analyze all the phases and quantum critical points in detail and give an accurate ground-state phase diagram. Finally, in Sec. ~\ref{CONCLUSIONS}, we provide a brief summary and discussion of the entire paper.

\section{MODEL AND METHODS}\label{MODEL AND METHODS}
We consider the half-filled Hubbard model on the SHO or BDS lattice (Fig.~\ref{fig1-Lattice}) with $D_{2}$ crystalline point-group symmetry. The Hamiltonian is given in the following,

\begin{equation} 
	\begin{aligned}
		H=&-\sum_{{\langle}i,j{\rangle}\sigma}t_{ij}(c^\dagger_{i\sigma}c_{j\sigma}+c^\dagger_{j\sigma}c_{i\sigma}) \\ &+\frac{U}{2}\sum_{i}(n_{i\uparrow}+n_{i\downarrow}-1)^2,
	\end{aligned} 
	\label{eq:1} 
\end{equation}
where ${\langle}i,j\rangle$ denotes the nearest-neighbor bonds, $c^\dagger_{i\sigma}$ and $c_{i\sigma}$ are creation and annihilation operators for fermions at site $i$ with spin ${\sigma}= \uparrow,\downarrow$, and $n_{i\sigma}  = c^\dagger_{i\sigma}c_{i\sigma}$ is the operator of electron occupation. In the biphenylene network, DFT calculation \cite{Ye_2023} shows that the hopping amplitudes within four-site plaquette are stronger than the other, therefore, we choose two kinds of hopping amplitudes in the DQMC calculations: $t_{ij} = t$ for the hopping within the four-site plaquette, and $t_{ij} = t'$ for the hopping on the ethylene bonds, as shown in Fig.~\ref{fig1-Lattice}(a). $U \geq 0$ denotes the strength of the on-site repulsion. 
On the SHO and BDS lattice, we set the nearest-neighbor bond $a=1$ as the length unit, the first Brillouin zone and high-symmetry points are present in Figs.~\ref{fig1-Lattice}(c-d).

In this paper, we employ the projective formalism of determinant quantum Monte Carlo (DQMC) \cite{Assaad2008}. 
The DQMC simulations were carried out based on the ALF package\cite{Assaad_2022ALF}.
The main idea of projective formalism of DQMC is that the nondegenerate ground-state of a many-body Hamiltonian can be projected from any known trial state that have nonzero overlap with the ground-state.
The ground-state expectation value of an observable can be obtained after projecting the trial wave-function
\begin{equation} 
	\begin{aligned}
	\langle \hat{O} \rangle = \frac{\langle \Psi_0 \mid \hat{O} \mid \Psi_0 \rangle}{\langle \Psi_0 \mid  \Psi_0 \rangle} = \lim\limits_{\Theta\rightarrow\infty} \frac{\langle \Psi^T \mid e^{-\Theta\hat{H}} \hat{O} e^{-\Theta\hat{H}} \mid \Psi^T \rangle}{\langle \Psi^T \mid e^{-2\Theta\hat{H}} \mid \Psi^T \rangle},
	\end{aligned} 
	\label{eq:3} 
\end{equation}
where $\Theta$ is the projection parameter. To make it be applicable by Monte Carlo calculation, we need to transform the quantum mechanics problem into multiple summation or integral of a classical problem. To realize that, we use Suzuki-Trotter decomposition \cite{trotter1959product, suzuki1991general} and introduce auxiliary field
\begin{equation} 
	\begin{aligned}
		e^{-\Theta\hat{H}} = (e^{-\Delta\tau\hat{H}_t} e^{-\Delta\tau\hat{H}_I})^M + \mathcal{O}(\Delta\tau^2),
	\end{aligned} 
	\label{eq:4}
\end{equation}
where $\Theta = \Delta\tau M$, $H_t$ is kinetic terms, and $H_I$ represents interaction term.  
We can decompose the two-body interaction term by introducing auxiliary field. Then the Hubbard-Stratonovitch transformation \cite{PhysRevLett.3.77} can be used to do that
\begin{equation} 
	\begin{aligned}
		e^{\Delta\tau\lambda\hat{O}^2} = \frac{1}{4}\sum_{l=\pm 1,\pm 2} \gamma(l)e^{\sqrt{\Delta\tau\lambda}\eta(l)\hat{O}} + \mathcal{O}(\lambda^4\Delta\tau^4).
	\end{aligned} 
	\label{eq:5} 
\end{equation}
 
As the $\Theta$ increases, and the $\Delta\tau$ decreases, the QMC results approach closer to the exact value. 
Considering the limitations of computational costs, we usually adopt a suitable $\Theta$ and $\Delta\tau$ in the simulation in order to get the accurate results.
Some details about the convergence tests for $\Theta$ and $\Delta\tau$ are shown in Appendix~\ref{test}.
In the following sections, we choose $\Delta\tau=0.1$ for calculating equal-time correlation function and $\Delta\tau=0.05$ for calculating time-displaced correlation function. 
In most cases, we set $\Theta=150$, while in the large $U$ region with larger gaps, $\Theta=100$ is also used to expedite the calculations.
And to study the physical quantities in the thermodynamic limit, we mainly use the linear system sizes $L = 2, 4, 6, 8, 10, 12$, to do the extrapolations. 
$L$ represents the number of unit cells along each primitive lattice vectors.
The total sites are equal to $6L^2$ where 6 is the number of sublattices within a unit cell.
Unless stated otherwise, all calculations in this paper are conducted using periodic boundary conditions (PBC).

 \begin{figure}[t]
	\centering
	\includegraphics[scale=0.397]{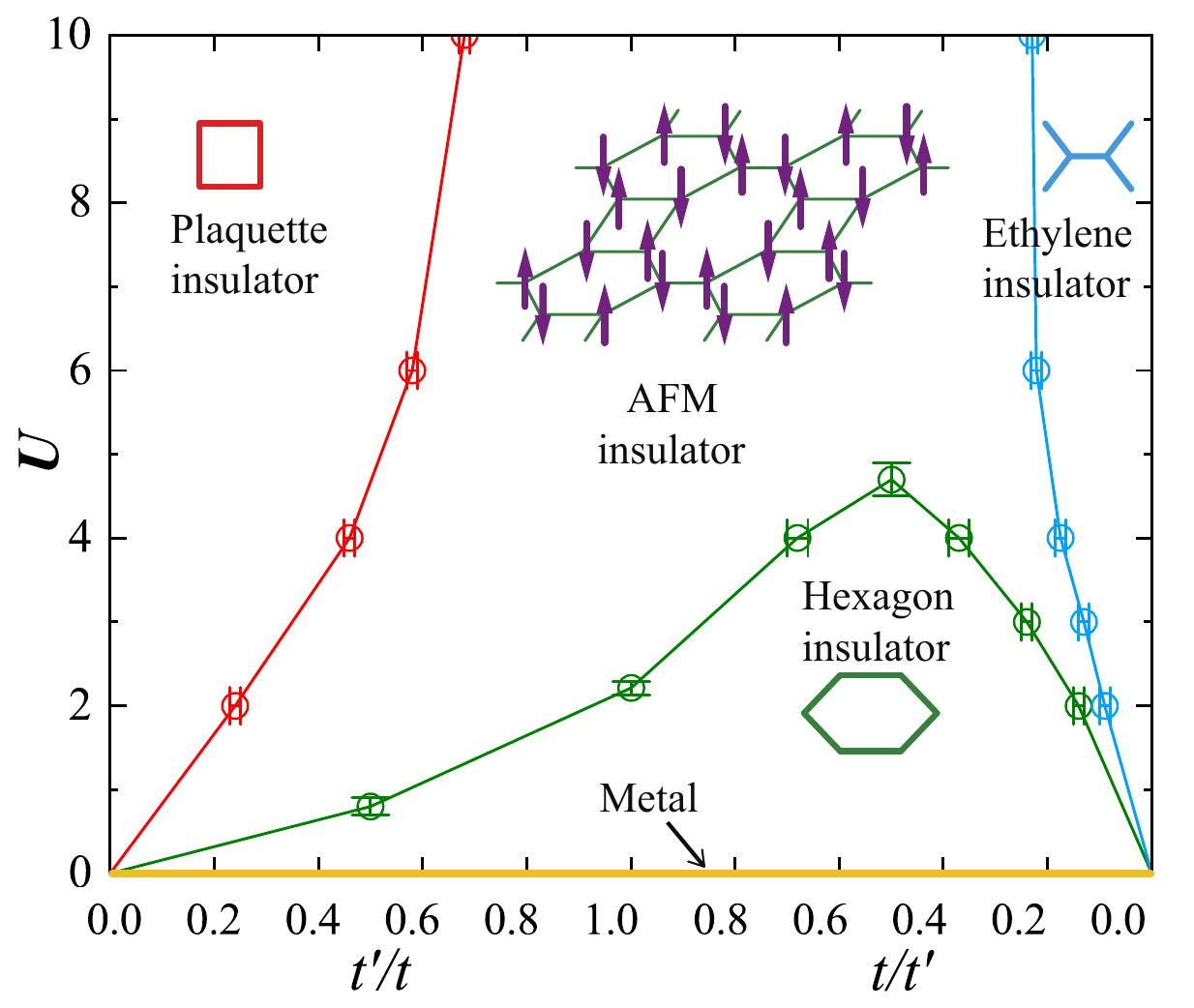}
	\caption{Phase diagram of the half-filled Hubbard model on the SHO lattice, the solid lines are the phase boundaries. 
		The plaquette insulator, ethylene insulator and hexagon insulator are separated by the N\'{e}el AFM Mott insulator. In the absence of on-site interaction (the yellow line with $U=0$), the entire system remains metallic no matter the ratio $t'/t$ or $t/t'$ is. According to the Hellmann-Feynman theorem, we exam the bond energy and its first-order derivative along the function path in Appendix~\ref{nophasetran}, revealing no phase transition signal and affirming that this region belongs to the same phase.}
	\label{fig2-phaseg} 
\end{figure}

In order to characterize possible magnetic order, we calculate the spin structure factor $S(\textbf{q}) = \frac{1}{N}\sum_{i,j}e^{i\textbf{q}\cdot (\textbf{r}_i - \textbf{r}_j)} \langle \textbf{S}_i \cdot\textbf{S}_j \rangle$ , where $\textbf{S}_i = \frac{1}{2} c^\dagger_{i} \pmb{\sigma} c_{i}$. For the sake of convenience, we show the $S(\textbf{q})$ by transforming SHO lattice into a BDS lattice [Fig.~\ref{fig1-Lattice}(b)] and then do the Fourier transform as the square lattice. 
The Bragg peak at $M^\prime(\pi,\pi)$ point is the signal of AFM ordering.
And we label the value of spin structure factor at this specific point as $S_{AF} = S(M')$

Apart from $S_{AF}$, to obtain the quantum critical points between AFM and nonmagnetic insulators quantitatively, we calculate the correlation ratio $R_{AF}$ of the structure factor which is defined as
\begin{equation} 
	\begin{aligned}
		R_{AF}(U, L) = 1 - \frac{S_{AF}(\textbf{Q} + d\textbf{q}, L)}{S_{AF}(\textbf{Q}, L)},
	\end{aligned} 
	\label{eq:7} 
\end{equation}
where $\textbf{Q}$ is the wave vector of the magnetic Bragg peak, and $\textbf{Q} + d\textbf{q}$ is the wave vector nearest to the peak position. This quantity is dimensionless. It tends to $1$ in the magnetically ordered phase and $0$ in the disordered phase, exhibits a size-independent behavior at the quantum critical point, and provides precise estimation of quantum critical points.

 \begin{figure}[!tb]
	\centering
	\includegraphics[scale=0.39]{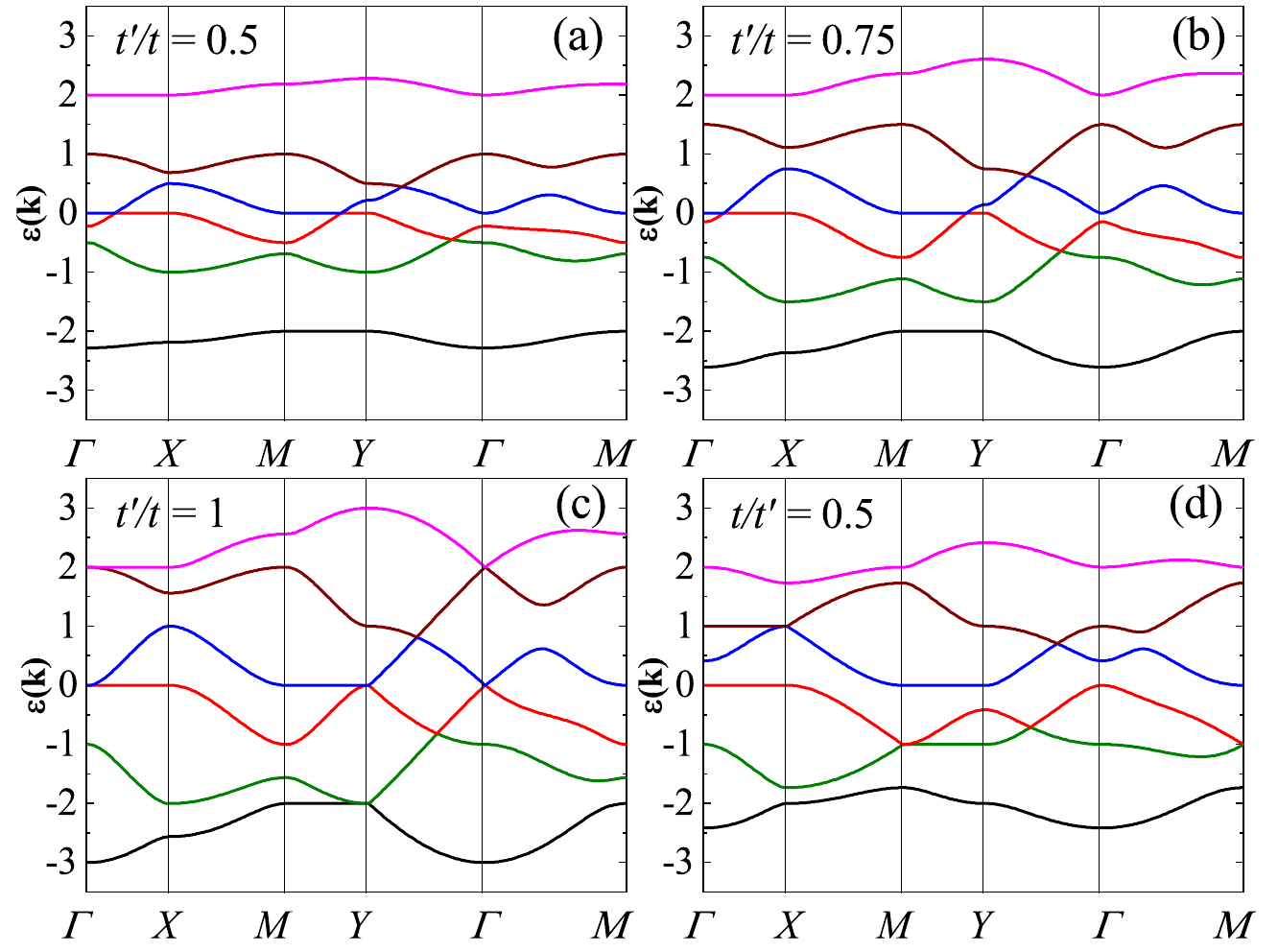}
	\caption{Non-interacting band structure of the SHO lattice along the high-symmetry path. In the region of $0 < t'/t \leq 1$, e.g. (a-c), there are some accidental band crossing points; in the region of $1 > t/t' > 0$, e.g. (d), two bands near the Fermi energy do not overlap directly.}
	\label{fig1} 
\end{figure}

In addition, to study the possible metal-insulator transition, we calculate the single-particle gap $\Delta_{sp}$, which can be extracted from the Matsubara  Green's function $G(\textbf{k},\tau) \propto exp(-\tau \Delta_{sp}(\textbf{k}))$ in the large imaginary time $\tau$. Similarly, the spin gaps can be obtained from the imaginary-time displaced spin-spin correlation functions $S_s(\textbf{k}, \tau) \propto exp(-\tau \Delta_s (\textbf{k}))$. 
Using these two gaps, we can identify whether some phases are gap or gapless. 
Furthermore, utilizing the maximum entropy (MaxEnt) method \cite{PhysRevB.57.10287, arxiv0403055} in the ALF package \cite{Assaad_2022ALF}, we can also extract spectral information from the aforementioned imaginary-time functions, where the correlation effects originated from Hubbard interaction can be clearly seen. 

\section{Numerical Results}\label{RESULT}

The full phase diagram obtained from DQMC simulations is shown in Fig.~\ref{fig2-phaseg}. Four different insulating phases exist: plaquette insulator, ethylene insulator, hexagon insulator and N\'{e}el AFM Mott insulator. The hexagon insulator is lying between AFM phase and the non-interacting metallic phase. 
Based on our results, the critical exponents between plaquette (ethylene) nonmagnetic insulating phase and AFM Mott insulating phase is very close to the 3D O(3) universality class\cite{PhysRevLett.61.2484, PhysRevLett.101.127202, doi:10.1143/JPSJ.66.2957, PhysRevB.65.014407, PhysRevB.79.014410, PhysRevB.99.174434, PhysRevB.73.014431, PhysRevB.65.144520}. 
However, for the phase transition between the hexagon insulating phase and the AFM phase, the values of the critical exponents deviate from the 3D O(3) universality class and closer to the chiral Heisenberg universality class using linear system sizes $L\leq12$.
We attribute this deviation from 3D O(3) universality class to the finite-size effect and the small single-particle around quantum critical points.
In the following three sections, we will explain how we obtained the phase diagram, and show the physical properties of these phases.

\subsection{Band structure at $U=0$}

Figure.~\ref{fig1} shows the non-interacting band structure of the SHO lattice. 
In the half-filling case, there is always metallic state with some flat dispersions along high-symmetry lines and some accidental crossing points at the Fermi level. The flat dispersions contribute to the divergent density of state at the Fermi level which is unstable to the local interactions.
In the region of $0 < t'/t < 1$, the bands around the Fermi level touch at a momentum point between $\varGamma$ and $X$, forming a type-II Dirac point \cite{soluyanov2015type, PhysRevLett.115.265304, PhysRevLett.119.026404, PhysRevLett.117.077202, PhysRevLett.117.086402, guan2017artificial, PhysRevB.108.144407}.
Within this region, the system maintains $D_{2}$ crystalline point-group characterized by four group elements and four one-dimensional inequivalent irreducible representations. 
Consequently, this band touching point can be classified as an accidental degeneracy \cite{Yang2014ClassificationOS, PhysRevB.85.195320, science1245085, PhysRevB88125427, PhysRevLett113027603}. 
In other words, it is nonrobust and can potentially be disrupted by small local interactions. 
As $t'/t$ increases, the intersection moves toward $\varGamma$. 
When $t'=t$, two Dirac points merge together at $\varGamma$ point with linear band touching along $\varGamma\text{-}Y$ direction and quadratic band touching along $\varGamma\text{-}X$ direction. 
In the region of $1 > t/t' > 0$, the bands do not overlap directly.
The introduction of Hubbard $U$ would push upper and lower bands away from Fermi energy.

\begin{figure}[t]
	\centering
	\includegraphics[scale=0.38]{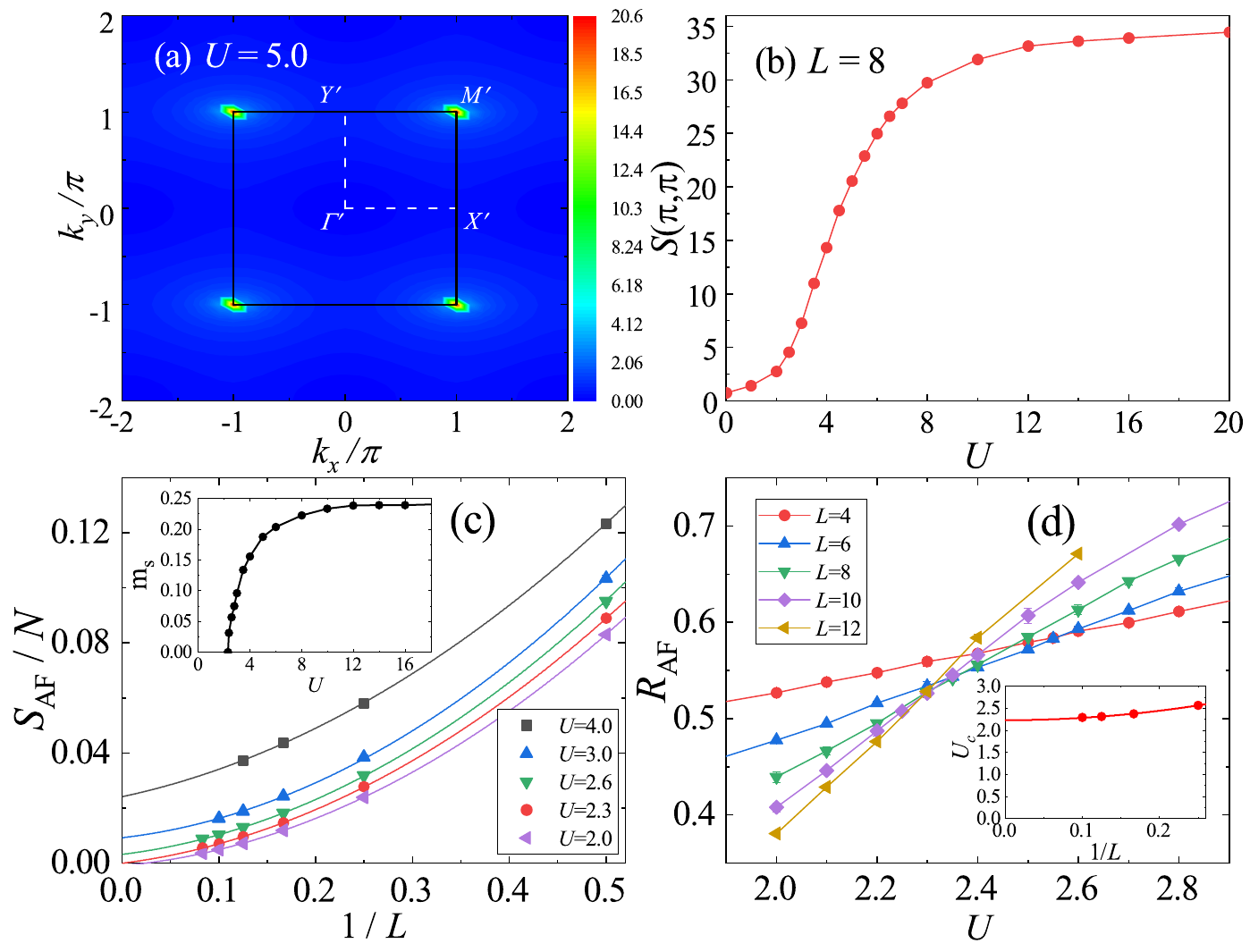}
	\caption{(a) The contour plots of static spin structure factor $S(\textbf{q})$ obtained with $L=8$, $t=t'=1$ and $U=5$. $S(\textbf{q})$ exhibits a sharp peak at $(\pi, \pi)$. (b) The value of $S(\pi,\pi)$ as a function of $U$. The linear system size is $L=8$. (c) Second-order polynomial extrapolation of $S_{AF}/N$ to TDL for $t=t'=1$. Inset shows the magnetic moment (square root of $\lim\limits_{N\to\infty}S_{AF}/N$) as a function of $U$. (d) Correlation ratio $R_{AF}$ as a function of interaction strength $U$ for $t=t'=1$. The intersection points indicate the quantum phase transition between disorder phase and magnetic ordered phase. Inset, the size dependence of $U_c(L)$ of $(L,L+2)$ crossings. The joint fits give $U_c \approx 2.23 $}
	\label{u5L8g}
\end{figure}

\subsection{Phase transition along $t'/t=1$}\label{t1}

What we are most concerned about is how the system changes after introducing the on-site Coulomb repulsion $U$. 
In this section, we will fix $t'=t=1$ to study the correlation effects induced by $U$.
When $U$ is large enough, due to the bipartite nature, the SHO lattice system often develops a long-range antiferromagnetic order.
To characterize this magnetic order, we show the static spin structure factor $S(\textbf{q})$ in Fig.~\ref{u5L8g}(a). At $U=5$, the static spin structure factor show sharp peak around $(\pi, \pi)$ point, which exhibits a N\'{e}el AFM Bragg peak. From Fig.~\ref{u5L8g}(b), it can be observed that the value of $S(\pi,\pi)$ is monotonically increasing with growing $U$, that indicates no other magnetic phase exists.
Here we want to mention that all the data we present in the paper contains error bars, and in most cases, the errors are very small and may hidden behind the data points.

Next, we want to investigate where the AFM order begins to emerge.
Figure~\ref{u5L8g}(c) shows the finite-size extrapolations of $S_{AF}/N$ with different on-site Coulomb repulsion $U$ to the thermodynamic limit (TDL). 
From the extrapolations, the AFM order occurs at $U\approx2.3$. 
The square root of extrapolated values $m_s=\sqrt{S_{AF}/N}$ as a function of $U$ are also shown in the inset of Fig.~\ref{u5L8g}(c). 
Above the $U_{c}$, the AFM order parameter $m_s$ exhibits a monotonic increasing behavior.
We also use a dimensionless quantity which is called correlation ratio $R_{AF}$ to identify the location of quantum critical point shown in Fig.~\ref{u5L8g}(d). The extrapolation of the joint of neighboring sizes shown in inset gives an more accurate estimation of the location of the quantum critical point $U_{c}\approx2.23$. 
The above results suggest that AFM order occurs at finite values of $U$ along $t=t'=1$ vertical line in the $t\text{-}t'\text{-}U$ plane.

\begin{figure}[t]
	\centering
	\includegraphics[scale=0.367]{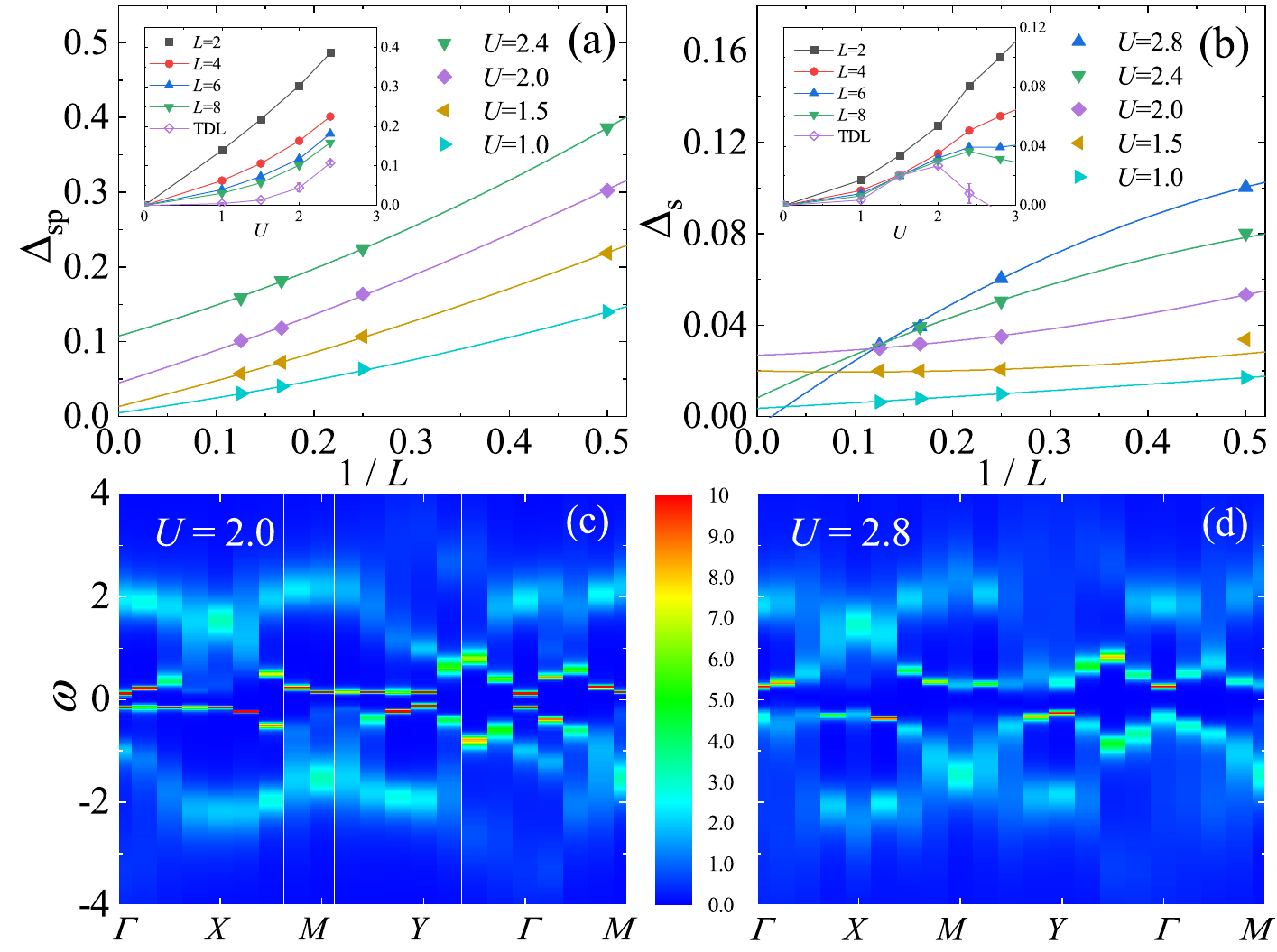}
	\caption{(a) Extrapolation of single particle gap $\Delta_{sp}$ to TDL by fixing $t=t'=1$. We use second-order polynomial functions to do the extrapolation. Inset, $\Delta_{sp}$ for $L=2,4,6,8$ and the extrapolated values (TDL), as functions of $U$. (b) The fitting of spin gap $\Delta_{s}$ with second-order polynomials in $1/L$. Inset, $\Delta_{s}$ for $L=2,4,6,8$ and the extrapolated values (TDL), as functions of $U$. (c) and (d) show the single-particle spectral function $A(\textbf{k},\omega)$ along the high symmetry path at $U = 2.0$ (hexagon insulating phase) and $U = 2.8$ (AFM Mott insulating phase), respectively. And the system size used here is $L=8$.}
	\label{t1gap}
\end{figure}

\begin{figure}[t]
	\centering
	\includegraphics[scale=0.37]{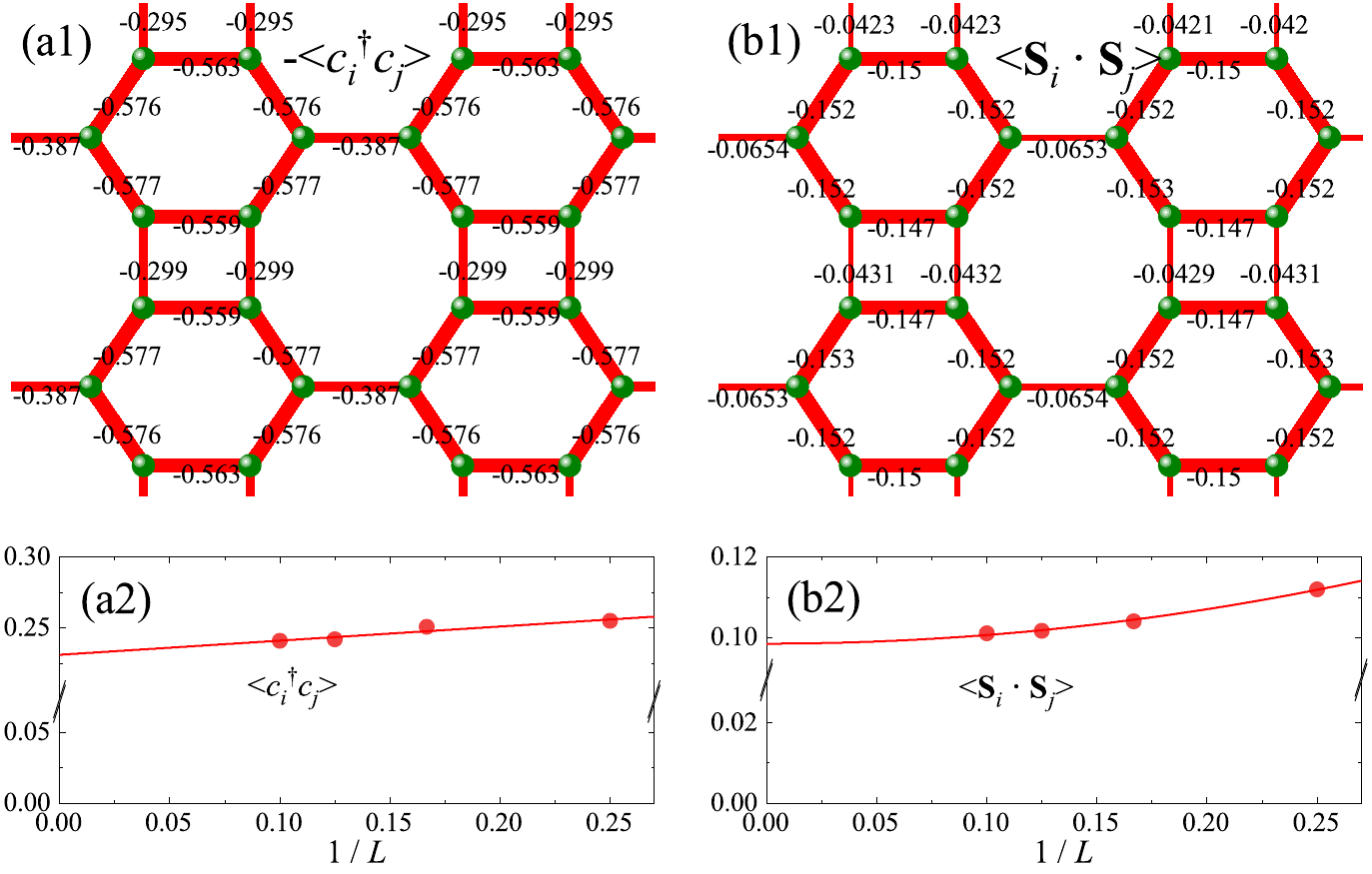}
	\caption{Nearest-neighbor (a1) kinetic energy $-\langle c^\dagger_{i}c_{j}\rangle$ and (b1) spin correlation $\langle \textbf{S}_i \cdot \textbf{S}_j \rangle$ at $t=t'=1$, $U=1.0$. The calculation were performed with $L=8$ lattice under open boundary condition. Here, we only show the middle four unit cells of them. The thickness of the lines indicate the relative magnitudes. And the numerical values are also shown on the bonds. (a2) and (b2) display the absolute difference of the corresponding values inside and between hexagons (The average value inside the hexagon minus the average value outside the hexagon). In the TDL, the correlations within the hexagon evidently dominate.}
	\label{T1U1citcjg}
\end{figure}

The appearance of the quantum critical point (QCP) at finite $U$ implies the presence of a non-magnetic phase in smaller $U$ region.
In the next step, we need to answer what the nonmagnetic phase is, and whether it is a metallic phase or an insulating phase. 
To further explore the nonmagnetic phases in the intermediate $U$ region, we calculate the single-particle gap $\Delta_{sp}$ from the time-displaced green function $G(\textbf{k},\tau)$, and obtain the spectral function $A(\textbf{k},\omega)$ via analytical continuation using the maximum entropy method \cite{PhysRevB.57.10287, arxiv0403055}. 
The single-particle gap is the minimum energy necessary to extract (add) one fermion from (to) the system, corresponds to the gap that can be observed in photoemission experiments and used to distinguish the metallic or insulating phase. In Fig.~\ref{t1gap}(a), we present the single-particle gap $\Delta_{sp}$ obtained from the DQMC simulations. The single-particle gap exhibits a continuous increasing with $U$ for all system sizes. Finite-size extrapolations of the available data points suggest a finite $\Delta_{sp}$ has already developed before the magnetic transition. 

To numerically find out whether the single-particle gap opens at $U=0$ or at finite $U$ (e.g. $0<U<1$) is technically challenging due to the extremely small gap when $U$ is small.
However, from analyzing the double occupancy $\langle n_{i\uparrow} n_{i\downarrow} \rangle$ in Appendix~\ref{nophasetran}, we do not see any signals of finite-$U$ metal-insulator-transition before the magnetic transition. We believe the low-$U$ phase is an nonmagnetic insulating phase. 
From the spectral function at $U=2.0$ shown in Fig.~\ref{t1gap}(c), before the AFM order emerges, the spectral function $A(\textbf{k},\omega)$ exhibits a noticeable gap at the Fermi level. The energy bands near Fermi level are still quasiparticle like with coherent peaks, while the other bands away from the Fermi level become incoherent. At $U=2.8$ within the AFM phase, the gap of the spectral function $A(\textbf{k},\omega)$ becomes larger, and the spectrum near Fermi level also becomes incoherent due to the strong correlation effect.
Therefore, we can conclude that the metallic phase is unstable to the interaction, and a very small on-site Coulomb repulsion $U$ will change it to an insulator. In other words, there is an intermediate insulating phase between non-interacting metallic phase ($U=0$) and AFM phase in the $t\text{-}t'\text{-}U$ phase diagram which is shown in Fig.~\ref{fig2-phaseg}.

Further details about the intermediate phase are obtained by examining the spin excitation gap $\Delta_s$ extracted from the time-displaced spin-spin correlation function. The AFM insulator, with a long-range magnetic order, breaks the $SU(2)$ spin rotational symmetry and has gapless Goldstone modes. Therefore, the spin gap should be zero in the TDL. For the noninteracting metallic phase, the spin gap is also zero. Figure.~\ref{t1gap}(b) shows the values of $\Delta_s$ obtained with different sizes and $U$, along with the extrapolations to the TDL. 
A finite value of $\Delta_s$ persists within the intermediate interaction regime, $0<U < U_c$. That means the intermediate nonmagnetic phase has not only nonzero single-particle gap but also nonzero spin gap. It is very similar to the valence-bond-solid-like phase\cite{10.1063/1.3518900, PhysRevLett.98.227202, PhysRevLett.123.157601}. 

\begin{figure}[t]
	\centering
	\includegraphics[scale=0.325]{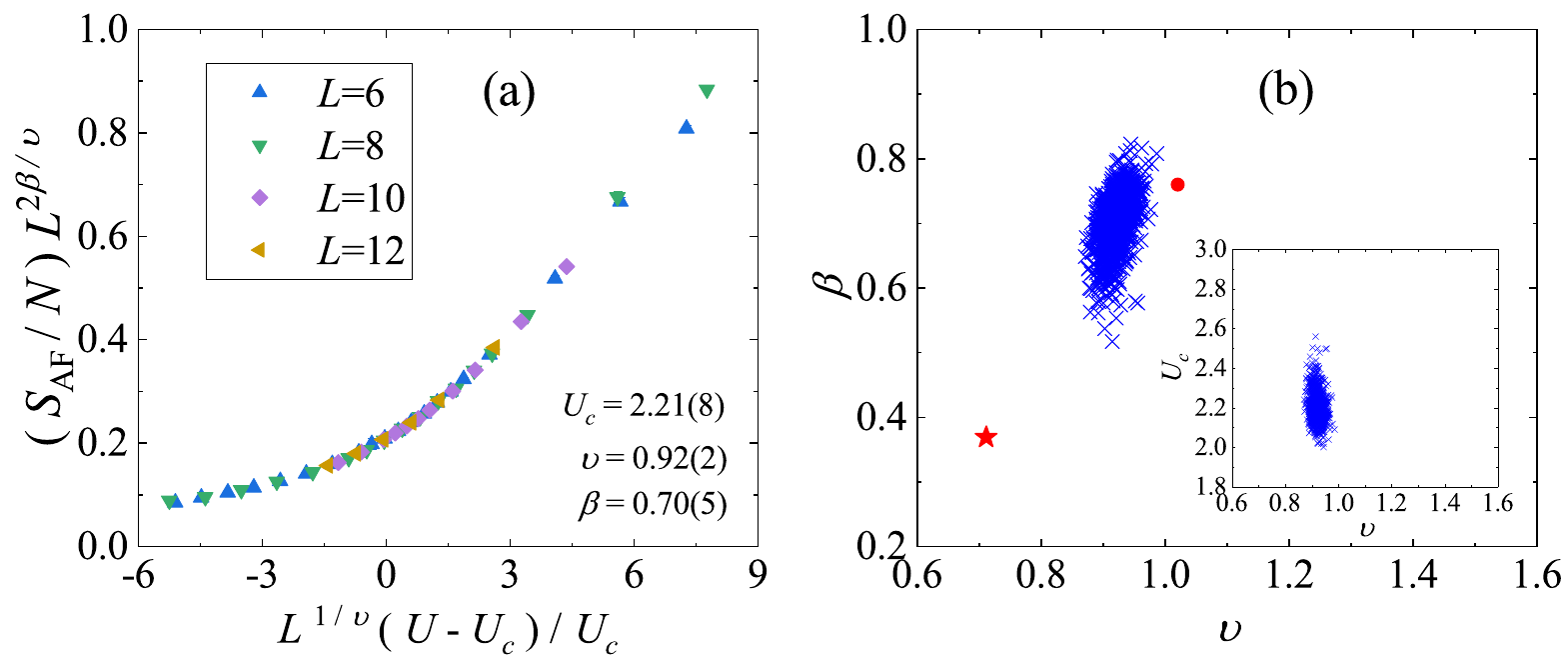}
	\caption{(a) Data-collapses of the $S_{AF}/N$ close to the critical point to obtain the critical exponents, along $t=t'=1$. The best parameters used for the collapse are $U_c = 2.21(8)$, $\nu=0.92(2)$ and $\beta=0.70(5)$. (b) Scattering plots of the optimal fitting parameters $U_c$, $\nu$ and $\beta$, after one thousand resampling. We calculate their mean value and standard deviation, and apply them to do the data-collapse shown in (a). The red dot represents the critical exponents of the chiral Heisenberg universality class \cite{S.Sorella_1992, PhysRevLett.97.146401, PhysRevB.80.075432, PhysRevX.3.031010, PhysRevX.6.011029,PhysRevB.106.075111, PhysRevD.96.096010}, while the red star denotes the critical exponents of the 3D O(3) universality class \cite{PhysRevLett.61.2484, PhysRevLett.101.127202, doi:10.1143/JPSJ.66.2957, PhysRevB.65.014407, PhysRevB.79.014410, PhysRevB.99.174434, PhysRevB.73.014431, PhysRevB.65.144520}.}
	\label{t1safv}
\end{figure}

To characterize whether the intermediate-$U$ phase is an VBS-like phase, we compute the nearest-neighbor kinetic energy $\langle c^\dagger_{i}c_{j}\rangle$ and spin-spin correlations $\langle \textbf{S}_i \cdot \textbf{S}_j \rangle$ under open boundary conditions (OBC) to break the translational symmetry, which are shown in Figs.~\ref{T1U1citcjg}(a1) and \ref{T1U1citcjg}(b1).
Under the influence of interactions $U$, both kinetic energy $\langle c^\dagger_{i}c_{j}\rangle$ and spin-spin correlations $\langle \textbf{S}_i \cdot \textbf{S}_j \rangle$ form noticeable hexagonal patterns. From the data in Figs.~\ref{T1U1citcjg}(a2) and \ref{T1U1citcjg}(b2), the kinetic energy $\langle c^\dagger_{i}c_{j}\rangle$ and spin-spin correlations $\langle \textbf{S}_i \cdot \textbf{S}_j \rangle$ within hexagons are larger than that ones between hexagons even in the thermodynamic limit. However, the hexagonal pattern phase does not break any translational or point group symmetry. We refer to this phase as a VBS-like hexagon insulating phase \cite{PhysRevB.64.144416, PhysRevLett.123.157601}.
It is worth noting that using $\langle c^\dagger_{i}c_{j}\rangle$ as the hopping amplitudes for calculating the band structure does not yield any Dirac points near the Fermi level. Consequently, it becomes implausible for the two Dirac points located at the $\varGamma$ point of the Brillouin zone to separate and shift away from $\varGamma$ upon entering the hexagon phase. Instead, a more likely result would be the direct opening of the single-particle gap at the $\varGamma$ point.

Finally, we want to elucidate the critical properties of the phase transition between the VBS-like hexagon phase and the AFM Mott insulating phase.
We collapse the $S_{AF}/N$ close to the phase transition with the finite-size scaling relation $S_{AF}/N = L^{-2\beta/\nu}f[L^{1/\nu}(U/U_c-1)]$.
The collapse data are shown in Fig.~\ref{t1safv}(a).
The finite-size scaling analysis is performed with a recently proposed method based on the Bayesian statistics \cite{PhysRevE.84.056704}. 
We follow the fitting process in \cite{PhysRevX.6.011029}, outlined below to obtain the optimal fitting parameters.
First, we prepare a data set $S_{AF}/N$ and introduce Gaussian noise based on the standard deviation of it. Second, we set appropriate initial values of the fitting parameters, $U_c=2.3$, $\nu = 0.9$, and $\beta=0.6$. Third, we perform the data-collapse analysis and fit the curve based on the Bayesian statistics \cite{PhysRevE.84.056704} to get the $\chi^2$. Finally, we adjust the parameters to closely align with the fitting curve (minimizing the $\chi^2$) to obtain the optimal parameters, $U_c$, $\nu$, and $\beta$.
By employing this method, we can reliably obtain the average and standard deviation of the fitting parameters.
After repeating the above procedure over a thousand times, we get the accurate phase transition point and critical exponents: $U_c = 2.21(8)$, $\nu=0.92(2)$ and $\beta=0.70(5)$ using four linear system sizes $L=6,8,10,12$. 
We project the three-dimensional scatter plot of the fitting parameters onto Fig.~\ref{t1safv}(b).
It is evident that the sampling data are closer to the red dot representing the critical exponents of the chiral Heisenberg universality class.

The universality class is determined by the symmetries inherent in the order parameter and the spatial dimensionality of the system. 
In the context of a two-dimensional (2D) system, a quantum phase transition, transitioning from a gapped valence-bond-solid-like insulating phase to the antiferromagnetic Mott insulating phase, typically aligns with the 3D O(3) universality class due to the spontaneous breaking of SU(2) symmetry. 
However, when the transition occurs within a weakly insulating phase characterized by a minute single-particle gap, and the size of the finite system remains insufficient relative to the correlation length within the fermionic sector, the fluctuations in the spin or magnetic order parameters might couple with low-energy fermionic excitations. This coupling can lead the universality class to deviate from conventional expectations\cite{PhysRevB.14.1165, PhysRevB.48.7183, MatthiasVojta_2003, Sachdev_1999}. 
In our specific scenario, near the quantum critical point, such as at $U=2.0$, both the single-particle gap and the spin gap are extremely small and of the same order, around $O(10^{-2})$, evident in the insets of Fig.~\ref{t1gap}(a) and Fig.~\ref{t1gap}(b). Even with $L=12$, the system size remains small in comparison to the fermionic correlation length ($\xi_e$), estimated to be approximately $O(10^{2})$ and roughly inversely proportional to the single-particle gap. 
However, conducting extensive DQMC simulations with considerably larger system sizes ($L\geq12$) would progressively emphasize the dominance of the spin sector as the system size increases. Eventually, this evolution would likely lead to the critical exponents converging toward the 3D O(3) universality class.
The critical exponents obtained from data collapse using ``small" system sizes indicate $\beta=0.70(5)$ and $\nu=0.92(2)$. These values diverge from the expected 3D O(3) values of $\beta=0.3689(3)$ and $\nu=0.7112(5)$\cite{PhysRevLett.61.2484, PhysRevLett.101.127202, doi:10.1143/JPSJ.66.2957, PhysRevB.65.014407, PhysRevB.79.014410, PhysRevB.99.174434, PhysRevB.73.014431, PhysRevB.65.144520}, but align more closely with the critical exponents of the chiral Heisenberg universality class $\beta=0.76(2)$ and $\nu=1.02(1)$ \cite{S.Sorella_1992, PhysRevLett.97.146401, PhysRevB.80.075432, PhysRevX.3.031010, PhysRevX.6.011029,PhysRevB.106.075111, PhysRevD.96.096010}. We attribute this phenomenon to the fluctuations of Dirac fermions near $\varGamma$ point and the coupling effect between the magnetic order parameter and very low-energy fermionic excitations.

\begin{figure}[t]
	\centering
	\includegraphics[scale=0.34]{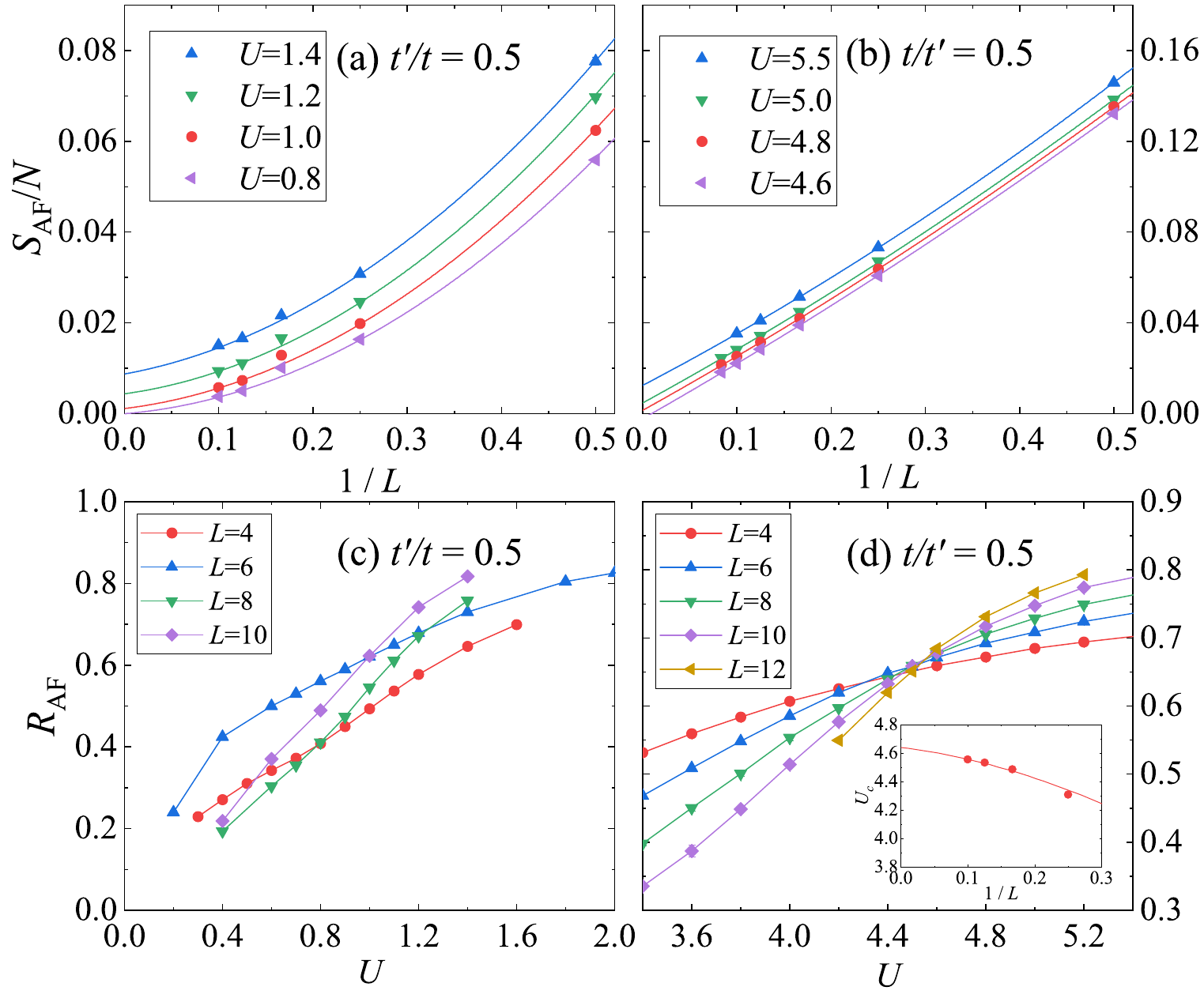}
	\caption{Extrapolation of $S_{AF}/N$ to TDL for (a) $t'/t=0.5$ and (b) $t/t'=0.5$. (c) and (d) are the corresponding correlation ratios $R_{AF}$ of the AFM order as a function of interaction strength $U$. Inset in (d) shows the variation of intersection points with different sizes. For $t/t'=0.5$, we obtain the QCP from two methods, and $U_c$ located between $4.6-4.8$. For $t'/t=0.5$, within the system sizes we could calculate, a concentrated intersection has not been observed. However, through the extrapolation in (a), we can roughly determine its $U_c\sim 0.8$ for $t'/t=0.5$.}
	\label{t5tp5saf}
\end{figure}

\subsection{Other QCPs}

\begin{figure}[t]
	\centering
	\includegraphics[scale=0.345]{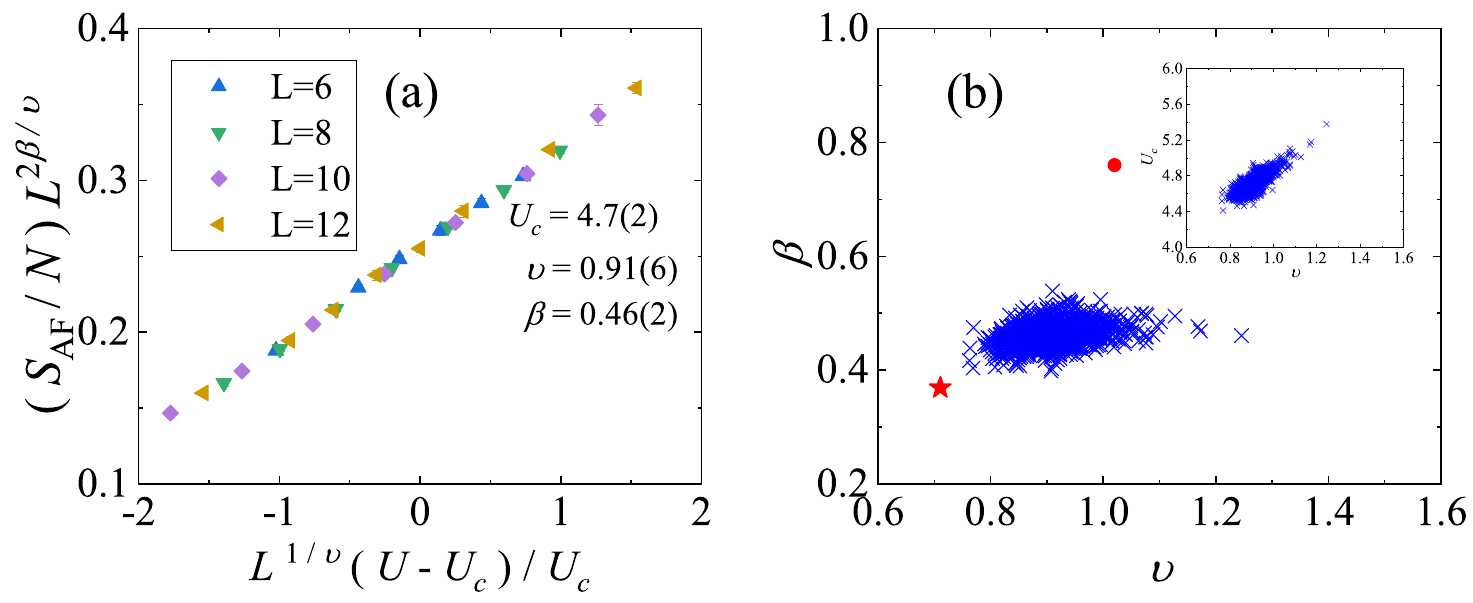}
	\caption{(a) Data-collapses of the $S_{AF}/N$ at $t/t'=0.5$ close to the critical point to obtain the critical exponents. The best parameters used for the collapse are $U_c = 4.7(2)$, $\nu=0.91(6)$ and $\beta=0.46(2)$. (b) Scattering plots of the optimal fitting parameters $U_c$, $\nu$ and $\beta$, after one thousand resampling. We calculate their mean values and standard deviations, and apply them to do the data-collapse shown in (a). The red dot represents the critical exponents of the chiral Heisenberg universality class \cite{S.Sorella_1992, PhysRevLett.97.146401, PhysRevB.80.075432, PhysRevX.3.031010, PhysRevX.6.011029,PhysRevB.106.075111, PhysRevD.96.096010}, while the red star denotes the critical exponents of the 3D O(3) universality class \cite{PhysRevLett.61.2484, PhysRevLett.101.127202, doi:10.1143/JPSJ.66.2957, PhysRevB.65.014407, PhysRevB.79.014410, PhysRevB.99.174434, PhysRevB.73.014431, PhysRevB.65.144520}.}
	\label{t05collapse}
\end{figure}

In this section, we will show how we map out the full phase diagram in the $t\text{-}t'\text{-}U$ plane.
Figure.~\ref{t5tp5saf} presents the finite-size scalings of magnetic orders and correlation ratio $R_{AF}$ to get the $U_c$ along $t'/t=0.5$ and $t/t'=0.5$.
For $t'/t=0.5$, the non-interacting band structure has a band touching point at 1/3 of the path from $\varGamma$ to $X$ as shown in Fig.~\ref{fig1}(a). Only when $L$ is an integer multiple of 6 this momentum point can be taken, causing the scaling behavior of $L=6$ system to deviate from that of other system sizes. Therefore, we only use $L=2,4,8,10$ to do the extrapolations.
Even worse, it is for this reason, we can not get a clear intersection in the correlation ratio $R_{AF}$. Anyway, we take the extrapolated results of magnetic structure factor $m_s^2$ shown in Fig.~\ref{t5tp5saf}(a) to estimate the quantum critical points. And we get $U_c=0.8(2)$ for $t'/t=0.5$. 
For $t/t'=0.5$, we can use both the extrapolation of ${m_s}^2$ and the crossing of $R_{AF}$ to get the QCP which is $U_c\sim4.7$.
To obtain more accurate QCP and critical exponents, we follow the same procedures as in the previous subsection~\ref{t1}. 
We obtain the following parameters: $U_c = 4.7(2)$, $\nu=0.91(6)$ and $\beta=0.46(2)$ shown in Fig.~\ref{t05collapse}, more closer to the critical exponents of 3D O(3) universality class than that at $t=t'=1$.
Around this QCP, the single-particle gap is larger than in the case where $t=t'=1$. Consequently, although there are still deviations from the precise critical exponents of the 3D O(3) universality class, the critical exponents here are now closer. This observation, when compared to the $t=t'=1$ case, strongly supports our explanation for the deviation of critical exponents.

\begin{figure}[t]
	\centering
	\includegraphics[scale=0.385]{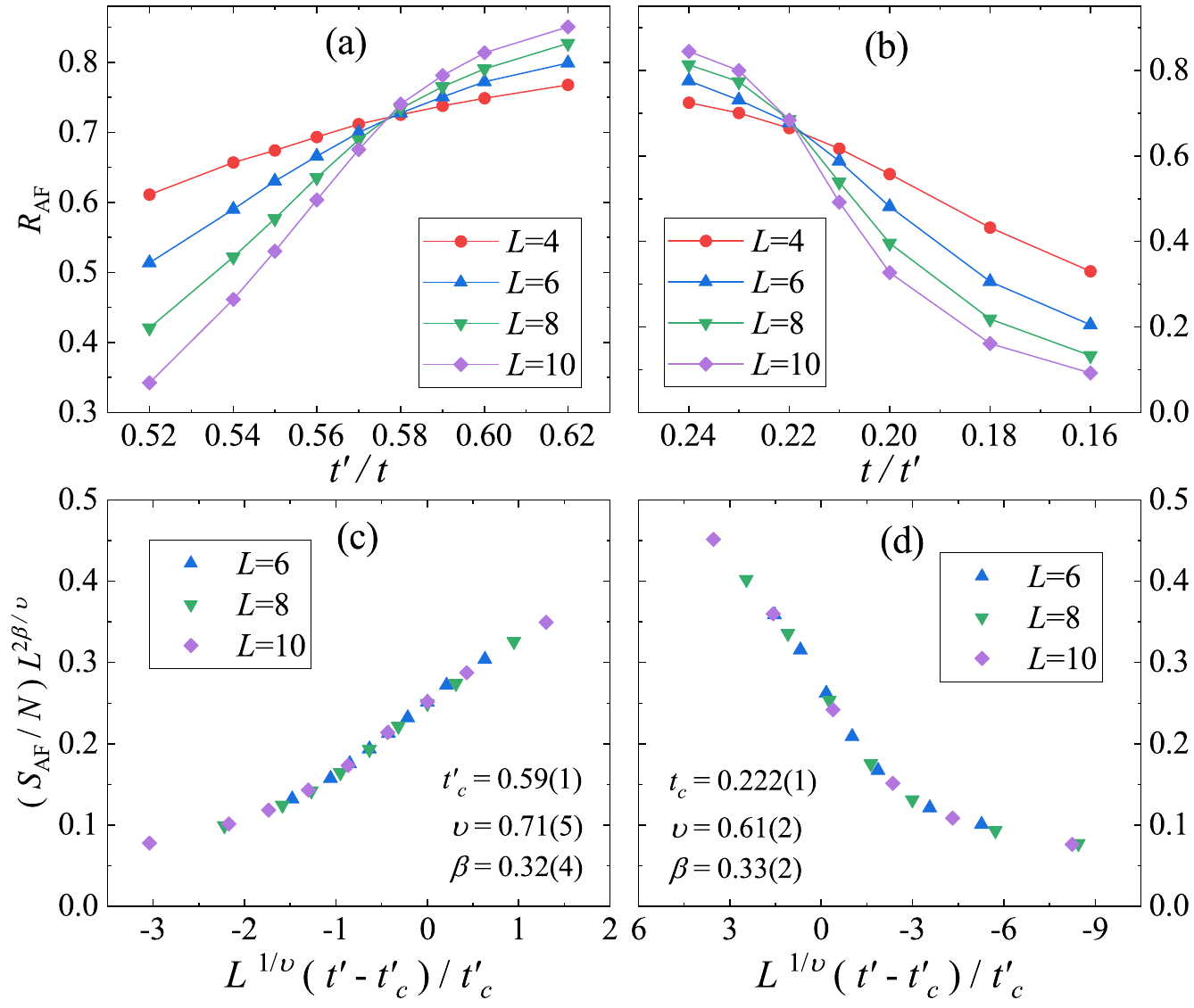}
	\caption{[(a) and (b)] Correlation ratio $R_{AF}$ close to the plaquette-AFM transition and AFM-ethylene transition. [(c) and (d)] Collapse of $S_{AF}/N$ close to the critical points. The parameters obtained form the Bayesian statistics are shown in the corresponding figure. When tuning $t'/t$, we fix $t=1$ and vary the value of $t’$ ; when tuning $t/t'$, we fix $t'=1$ and vary $t$.}
	\label{U6ARAF}
\end{figure}

In addition to the hexagon insulating phase and AFM phase mentioned earlier, there are other two insulating phases named plaquette insulator and ethylene insulator in the phase diagram. These two insulating phases are nonmagnetic phases and have their own decoupled limits.
Then, we want to get the QCPs between the AFM phase and these two phases. The plaquette VBS-like insulator is a fully gapped state which is adiabatically connected to the decoupled plaquette limit with the direct product of each plaquette singlets as ground state. Therefore the QCP between plaquette insulator and AFM ordered phase fits the formalism of classical critical point, and these QCP are called conventional QCP \cite{S0217979212300071}. 
It exists in many dimerized models \cite{PhysRevLett.101.127202, doi:10.1143/JPSJ.66.2957, PhysRevB.65.014407, PhysRevB.79.014410,  PhysRevB.99.174434, PhysRevB.73.014431} that have been thoroughly studied. 
The phase transition belongs to the 3D O(3) universality class \cite{PhysRevB.65.144520}. The ethylene insulator exhibits the same behavior. 
To get the critical points and critical exponents, we collapse the $S_{AF}/N$ with the finite-size scaling relation $S_{AF}/N = L^{-2\beta/\nu}f[L^{1/\nu}(t/t_c-1)]$ similar to that in Sec.~\ref{t1}. When $t'$ controls the phase transition, $t$ is replaced by $t'$. As shown in the Fig.~\ref{U6ARAF}(a) and (b), taking $U=6$ as an example, different sizes have nearly the same intersection points. The data collapse shown in Figs.~\ref{U6ARAF}(c) and \ref{U6ARAF}(d) obtain the accurate critical points and critical exponents. We show their numerics in the lower of the figures.
These two sets of critical exponents closely resemble the 3D O(3) critical exponents, which demonstrates that both phase transitions belong to the 3D O(3) universality class. 
Interestingly, the critical exponent $\nu$, which characterizes the correlation length, exhibits value at the QCP between the plaquette insulator and the AFM insulator that are closer to the 3D O(3) model compared to those obtained from the QCP between the ethylene insulator and AFM insulator. This observation may attributed to the fact that at $U=6.0$, the gaps of the plaquette insulator are larger than that of the ethylene insulator.

\begin{figure}[t]
	\centering
	\includegraphics[scale=0.4]{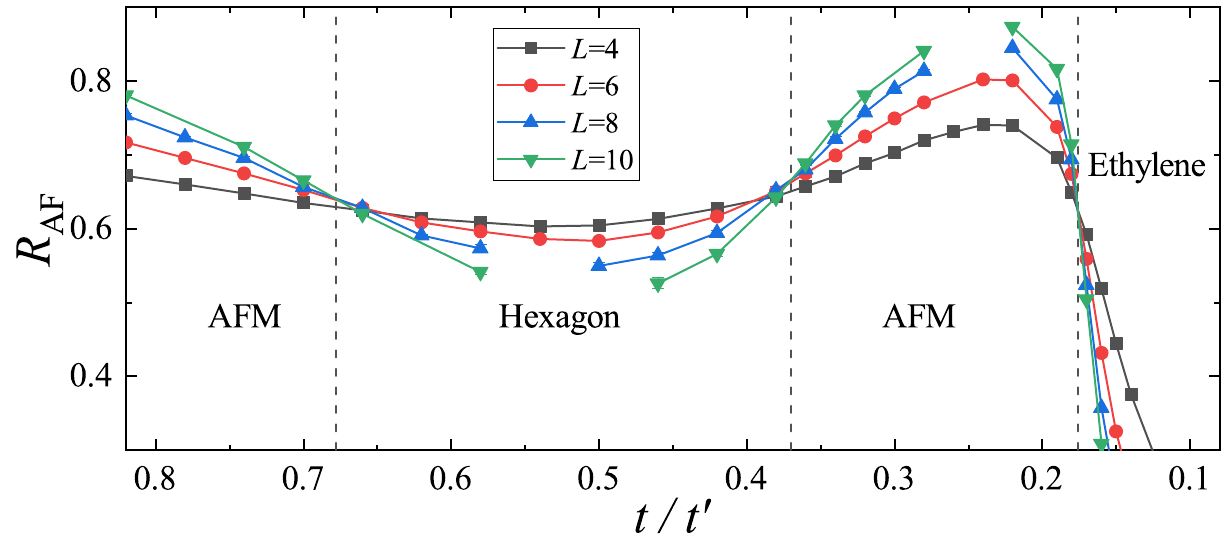}
	\caption{Correlation ratios $R_{AF}$ of the AFM order as a function of $t/t'$ at $U=4$, the protruding region represents the AFM phase sandwiched between two VBS-like insulating phases in the lower-right corner of the phase diagram.}
	\label{U4HRAFG}
\end{figure}

In the lower-right corner of the phase diagram shown in Fig.~\ref{fig2-phaseg}, hexagon insulating phase and ethylene insulating phase are very close. 
To determine whether there is a direct phase transition between them or if there is an intermediate AFM phase, we calculate the $R_{AF}$ as a function of $t/t'$. Taking $U=4.0$ as an example, from the data in Fig.~\ref{U4HRAFG}, we observe that there is an AFM phase sandwiched between the hexagon insulator and the ethylene insulator.
The AFM phase region decreases as $U$ reduces from $4.0$ to $3.0$ and further to 2.0.
Supplemented by the contour plot of $S_{AF}/N$ using a linear system size $L=4$ and the spin gaps of two decoupled limits ($t'/t=0$ and $t/t'=0$), as shown in Fig.~\ref{phaseDiagramL4}(a) of Appendix~\ref{L4Phase}.

 %Based on the contour plot of $S_{AF}/N$ using a system size of $L=4$ as shown in Fig.~\ref{phaseDiagramL4}(a) of Appendix~\ref{L4Phase}, we can reasonably conjecture that this AFM phase region will vanish in the $U=0$ limit.

\section{CONCLUSIONS}\label{CONCLUSIONS}
In summary, we have mapped out the ground-state phase diagram of the Hubbard model on the SHO lattice at half filling with projective formalism of DQMC simulation. Based on the numerical results, we observe a AFM phase surrounded by three VBS-like insulators. The phase transition between AFM and plaquette (ethylene) insulator belongs to 3D O(3) universality class.
In addition, we have computed the critical exponents at the QCP between the AFM phase and the VBS-like hexagon phase.
It seems that this QCP deviates from the 3D O(3) universality class. we attribute this deviation to the small single-particle gap with strong charge fluctuation in the hexagon phase which corresponds to finite-size effect. The critical exponents are expected to fall into the 3D O(3) universality class in the thermodynamic limit. 
We hope our systematic numerical results will motivate further experimental investigations of the correlation effects on the SHO lattice in transition metal oxides or cold atoms.

The hexagon insulating phase shares similarities with the plaquette and ethylene insulating phases. Therefore we speculate that it might adiabatically connect to the decoupled hexagon limit, where only nearest-neighbor hopping within hexagon exists.
To confirm this, we can change the ratio of the nearest-neighbor hopping within hexagons ($t$) to the nearest-neighbor hopping between hexagons ($t'$) to explore the possible adiabatic connection between hexagon phase at $t=t'=1$ and the decoupled limit.
Further details will be explored in our future research.

For the biphenylene network, first principle calculations show the band structure near the Fermi level does not have flat bands \cite{doi:10.1021/nn100758h, GE201697, Ye_2023}. To fit this band structure, Ref.\cite{Ye_2023} shows the tight binding model needs 15 different hopping energies to achieve a more accurate fit. Our simpler model is not accure enough to study the correlation effect on the biphenylene material. Unfortunately, we cannot apply DQMC to study the fitting model due to Fermi sign problem. However, other methods like density matrix renormalization group and tensor network are suitable there, which we also leave for future study.

\section{ACKNOWLEDGMENTS}\label{ACKNOWLEDGMENTS}

We would like to thank Shuai Yin, Zhongbo Yan and Muwei Wu for the helpful discussions.
We are also grateful to F. F. Assaad for providing the open-source software package and valuable suggestions on code modifications.
This project is supported by NKRDPC-2022YFA1402802, NSFC-11804401, NSFC-92165204, NSFC-11974432, Leading Talent Program of Guangdong Special Projects (201626003), and Guangzhou Basic and Applied Basic Research Foundation (202201011569).
The calculations reported were performed on resources provided by the Guangdong Provincial Key Laboratory of Magnetoelectric Physics and Devices, No. 2022B1212010008.

\appendix

\section{Imaginary-time displaced Green's function and imaginary-time displaced spin-spin correlation function}

\begin{figure}[h]
	\centering
	\includegraphics[scale=0.44]{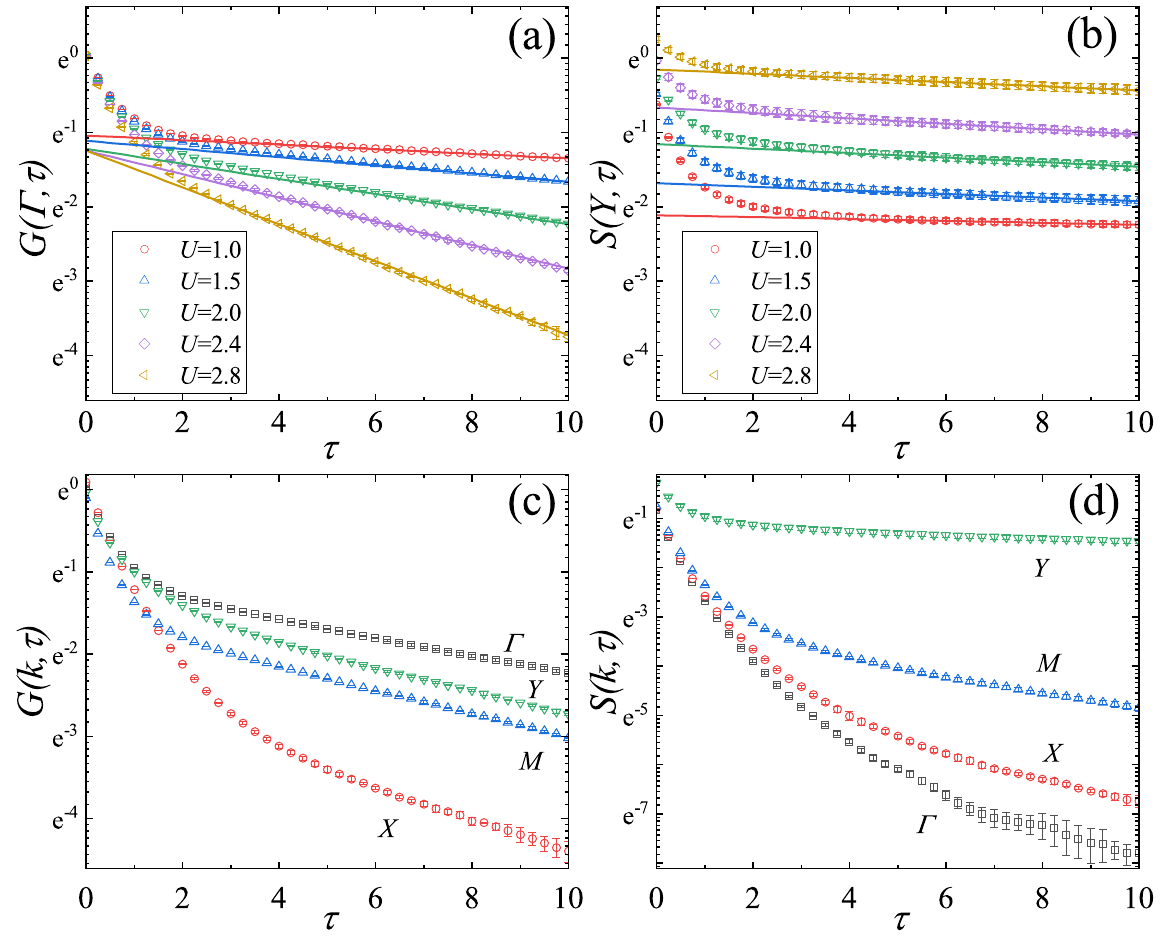}
	\caption{Semilogarithmic plots of (a) imaginary-time displaced Green's function and (b) imaginary-time displaced spin-spin correlation function with $t/t'=1$, $L=8$ for different $U$. Single-particle gap and spin excitation gap are obtained by linearly fitting the tail of them. (c) and (d) correspond to the $G(\textbf{k},\tau)$ and $S(\textbf{k},\tau)$ in some high-symmetry momentum points with the same linear system size $L=8$. We can see that the lowest single-particle gap is at the $\varGamma$ point, while the lowest spin gap at the $Y$ point.}
	\label{T1L8SPGV} 
\end{figure}

Single-particle gap $\Delta_{sp}(\textbf{k})$ can be extracted from the imaginary-time displaced Green's function by $G(\textbf{k},\tau) \propto exp(-\tau\Delta_{sp}(\textbf{k}))$, corresponding to the difference between one particle excitation energy and chemical potential $\mu$. 
In this system, with half-filling electrons, the chemical potential $\mu$ is zero. 
Similarly, spin excitation gap can be obtained from the imaginary-time displaced spin-spin correlation function by $S(\textbf{k},\tau) \propto exp(-\tau\Delta_{s}(\textbf{k}))$. 
Figures.~\ref{T1L8SPGV}(a) and \ref{T1L8SPGV}(b) display the imaginary-time displaced Green's function and imaginary-time displaced spin-spin correlation function for different $U$, which can be used to extract the single-particle gap and spin excitation gap shown in Figs.~\ref{t1gap}(a) and \ref{t1gap}(b) of main text. Figs.~\ref{T1L8SPGV}(c) and \ref{T1L8SPGV}(d) reveal that the minimum value of single-particle gap occurs at $\varGamma(0,0)$, while the minimum value of spin gap appears at $Y(0,\pi)$.

\section{Phase transition within hexagon insulating phase}\label{nophasetran}

\begin{figure}[h]
	\centering
	\includegraphics[scale=0.44]{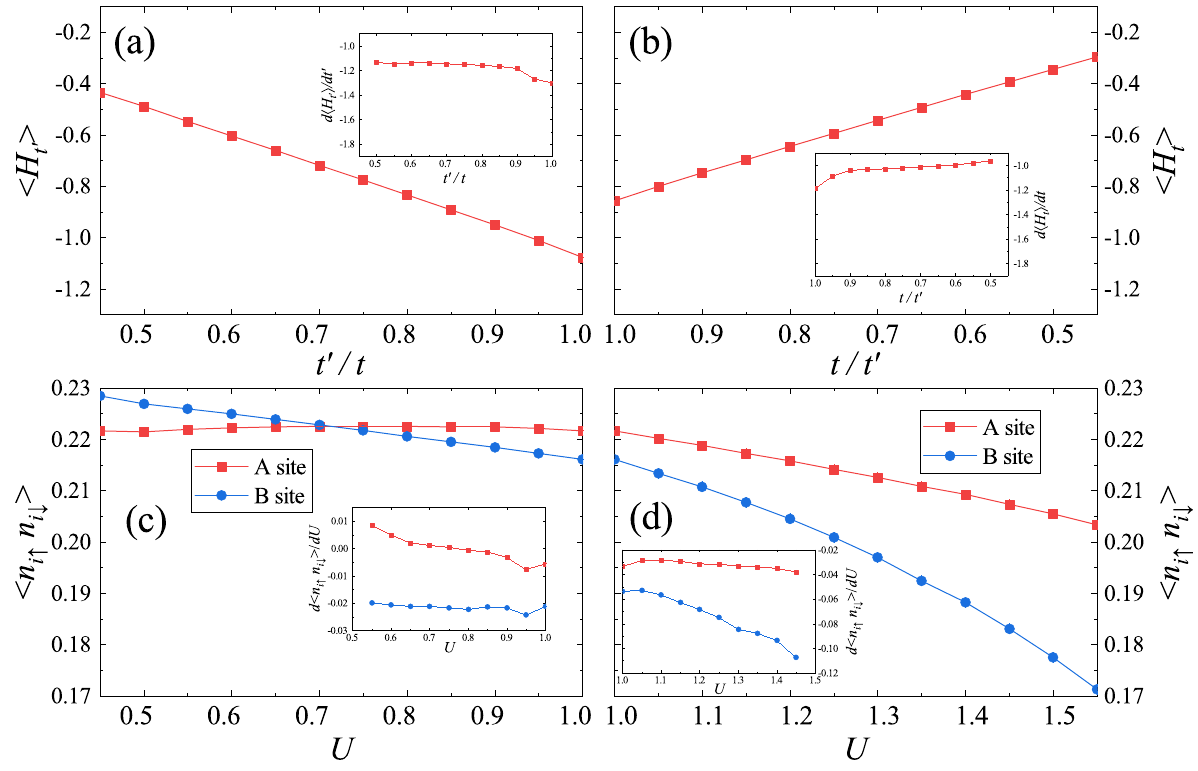}
	\caption{(a) $t'$ bond energy $\langle H_{t'} \rangle$ follow the function path of $U=t'/t$. (b) $t$ bond energy $\langle H_{t} \rangle$ follow the function path of $U=2-t/t'$. (c) The double occupancy $\langle n_{i\uparrow} n_{i\downarrow} \rangle$ of A site and B site on the same path as (a). (d) The double occupancy $\langle n_{i\uparrow} n_{i\downarrow} \rangle$ of A site and B site on the same path as (b). Insets correspond to the first-order derivative. The calculations above are conducted with $L=6$. }
	\label{Ettp} 
\end{figure}

To verify whether the nonmagnetic region below AFM phase in the phase diagram shown in Fig.~\ref{fig2-phaseg} is the same phase, we exam the bond energy on the red $t$ and light blue $t'$ bonds shown in Fig.~\ref{fig1-Lattice} (respectively named as $\langle H_{t} \rangle$ and $\langle H_{t'} \rangle$), and double occupancy along the function path of
 $$U=
 \begin{cases}
 	t'/t& 0.5<t'/t\leq1\\
 	2-t/t'& 1>t/t'>0.5.
 \end{cases}$$
As shown in Fig.~\ref{Ettp}, The absence of peaks or discontinuities in the first derivative results indicate that there is no quantum phase transition along this path.
Therefore we have confirmed that all the non-magnetic regions below the AFM phase belong to the same phase.

\begin{figure}[h]
	\centering
	\includegraphics[scale=0.44]{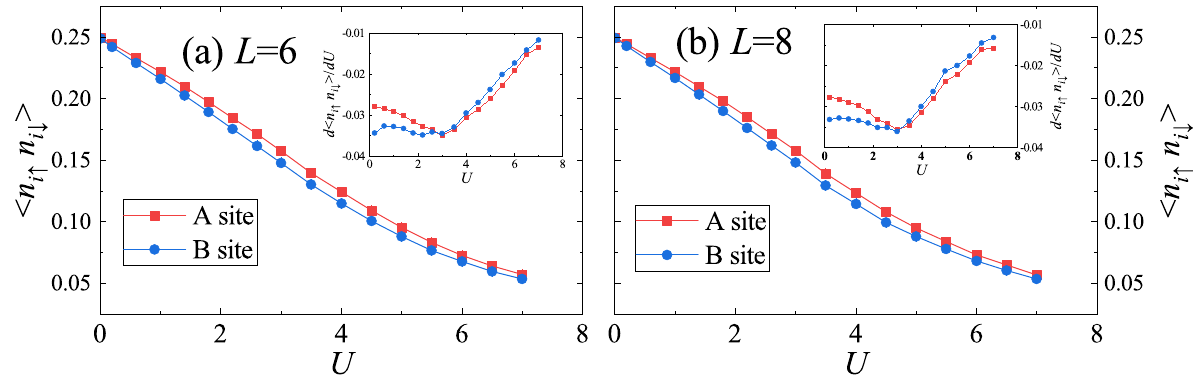}
	\caption{The double occupation $\langle n_{i\uparrow} n_{i\downarrow} \rangle$ of A site and B site as a function of $U$, at $t=t'=1$. 
		%C, E, and F site are topologically consistent with B, and D is consistent with A. 
		Insets are the first-order derivative of double occupancy.}
	\label{dblo} 
\end{figure}

In the main text, we observe a pronounced single-particle gap at $U\geq1$, $t=t'=1$.
It is difficult to find out whether the single-particle gap opens at $U=0$ or finite $1> U\geq0$. So that we show the double occupancy $\langle n_{i\uparrow} n_{i\downarrow} \rangle$ and its first-order derivative in Fig.~\ref{dblo}. 
In the range of $U\in\left[0-1\right)$, double occupancy $\langle n_{i\uparrow} n_{i\downarrow} \rangle$ monotonously decrease, with no discontinuity when $1> U >0$. This is also reflected in the first derivative. Except for the magnetic transition peak $U_c=2.21(8)$, there is no peak at $1> U >0$. Hence, we conclude that the low-$U$ phase is an nonmagnetic insulating phase.

\begin{figure}[t]
	\centering
	\includegraphics[scale=0.44]{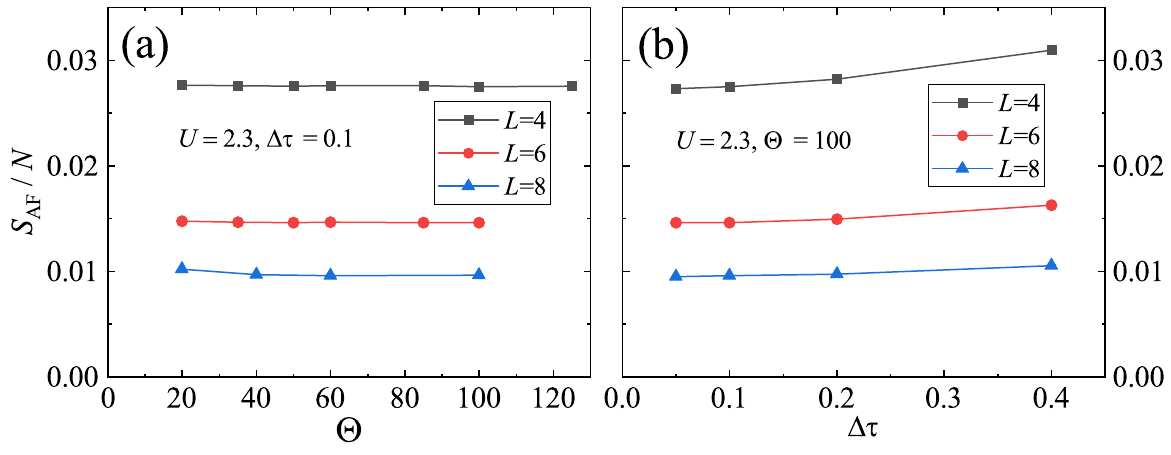}
	\caption{The AFM structure factor $S_{AF}/N$ are plotted with respect to different (a) $\Theta$ and (b) $\Delta\tau$, at $U=2.3$ and $t=t'=1$.}
	\label{testplot} 
\end{figure}

\section{The choices of projection time $\Theta$ and discrete time slice $\Delta\tau$}\label{test}

In this section, we provide convergence tests for projection time $\Theta$ and discrete time slice $\Delta\tau$ along $t'/t=1$.
For projection time $\Theta$, we test its convergence near the phase transition point at $U=2.3$ for different sizes. As shown in Fig.~\ref{testplot}(a), $\Theta\geq40$ is able to get the covergent structure factors for the $L=4,6,8$ system sizes. However, we choose $\Theta=100$. 
We also conduct the similar test on $\Delta\tau$, and the data are presented in Fig.~\ref{testplot}(b). 
$\Delta\tau = 0.1$ is already suitable for calculating the static correlation functions. For time-displaced green functions, we use $\Delta\tau=0.05$ to ensure longer projection time and longer tail to fit the gaps.

\begin{figure}[t]
	\centering
	\includegraphics[scale=0.38]{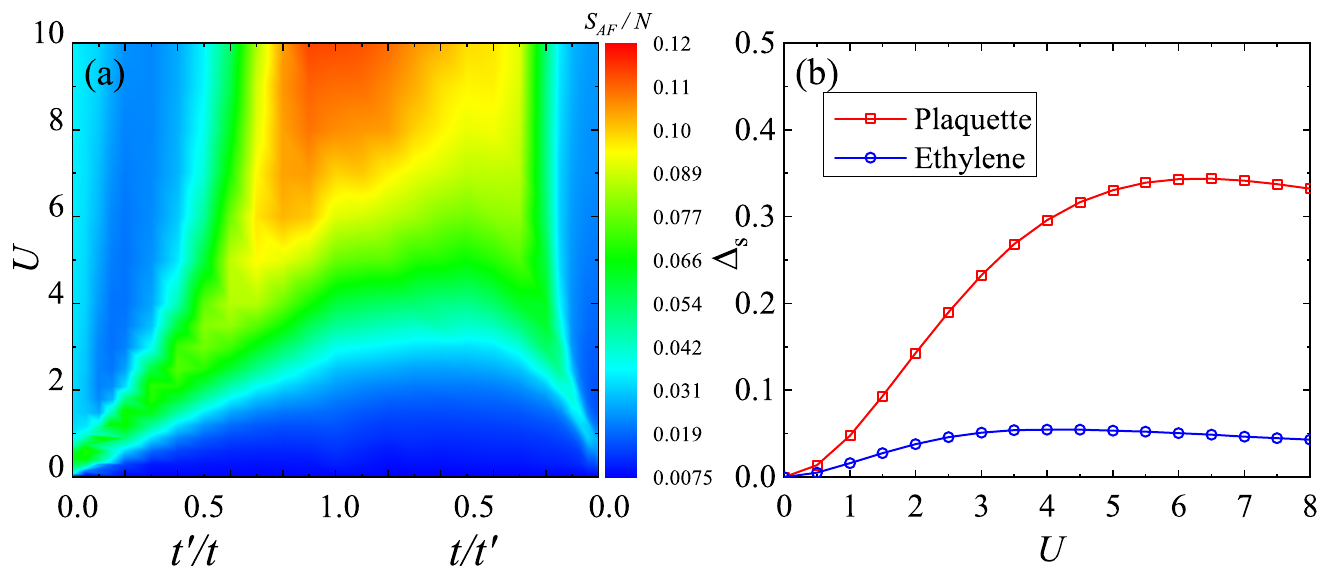}
	\caption{(a) The contour plot of AFM structure factor $S_{AF}/N$ for a smaller size $L=4$. (b) The spin gap $\Delta_s$ of one Plaquette or Ethylene calculated by exact diagonalization (ED).}
	\label{phaseDiagramL4} 
\end{figure}

\section{Reductions of the AFM region at two lower corners}\label{L4Phase}

To estimate the phase boundaries at lower left and lower right corners of phase diagram shown in Fig.~\ref{fig2-phaseg}. We computed the AFM structure factor for a smaller system size, $L=4$, spanning the entire phase diagram, depicted in Fig.~\ref{phaseDiagramL4}(a).
The green regions at low $U$ values indicate weak long-range antiferromagnetic order. These green regions appear to extend towards the limits of $U\rightarrow0$, $t'/t\rightarrow0$, and $U\rightarrow0, t/t'\rightarrow0$. From this observation, we can reasonably speculate that the AFM phase region persists for small $U$ values but disappears in the $U=0$ limit at these two limits.

Additional evidence supporting these viewpoints is derived from analyses of two decoupled limits. In the limits of $t'/t=0$ and $t/t'=0$, based on calculations of the lowest excitation gap (triplet gap) with four or six lattice sites, finite energy gaps are observed at $U>0$, as can be seen in Fig.~\ref{phaseDiagramL4}(b). These gaps can facilitate the extension of the plaquette phase or ethylene phase to finite values of $t'/t$ or $t/t'$ while keeping $U$ fixed at a finite value.

\bibliography{reference}

\begin{thebibliography}{64}%
\makeatletter
\providecommand \@ifxundefined [1]{%
 \@ifx{#1\undefined}
}%
\providecommand \@ifnum [1]{%
 \ifnum #1\expandafter \@firstoftwo
 \else \expandafter \@secondoftwo
 \fi
}%
\providecommand \@ifx [1]{%
 \ifx #1\expandafter \@firstoftwo
 \else \expandafter \@secondoftwo
 \fi
}%
\providecommand \natexlab [1]{#1}%
\providecommand \enquote  [1]{``#1''}%
\providecommand \bibnamefont  [1]{#1}%
\providecommand \bibfnamefont [1]{#1}%
\providecommand \citenamefont [1]{#1}%
\providecommand \href@noop [0]{\@secondoftwo}%
\providecommand \href [0]{\begingroup \@sanitize@url \@href}%
\providecommand \@href[1]{\@@startlink{#1}\@@href}%
\providecommand \@@href[1]{\endgroup#1\@@endlink}%
\providecommand \@sanitize@url [0]{\catcode `\\12\catcode `\$12\catcode
  `\&12\catcode `\#12\catcode `\^12\catcode `\_12\catcode `\%12\relax}%
\providecommand \@@startlink[1]{}%
\providecommand \@@endlink[0]{}%
\providecommand \url  [0]{\begingroup\@sanitize@url \@url }%
\providecommand \@url [1]{\endgroup\@href {#1}{\urlprefix }}%
\providecommand \urlprefix  [0]{URL }%
\providecommand \Eprint [0]{\href }%
\providecommand \doibase [0]{https://doi.org/}%
\providecommand \selectlanguage [0]{\@gobble}%
\providecommand \bibinfo  [0]{\@secondoftwo}%
\providecommand \bibfield  [0]{\@secondoftwo}%
\providecommand \translation [1]{[#1]}%
\providecommand \BibitemOpen [0]{}%
\providecommand \bibitemStop [0]{}%
\providecommand \bibitemNoStop [0]{.\EOS\space}%
\providecommand \EOS [0]{\spacefactor3000\relax}%
\providecommand \BibitemShut  [1]{\csname bibitem#1\endcsname}%
\let\auto@bib@innerbib\@empty
%</preamble>
\bibitem [{\citenamefont {Novoselov}\ \emph {et~al.}(2004)\citenamefont
  {Novoselov}, \citenamefont {Geim}, \citenamefont {Morozov}, \citenamefont
  {Jiang}, \citenamefont {Zhang}, \citenamefont {Dubonos}, \citenamefont
  {Grigorieva},\ and\ \citenamefont {Firsov}}]{doi:10.1126/science.1102896}%
  \BibitemOpen
  \bibfield  {author} {\bibinfo {author} {\bibfnamefont {K.~S.}\ \bibnamefont
  {Novoselov}}, \bibinfo {author} {\bibfnamefont {A.~K.}\ \bibnamefont {Geim}},
  \bibinfo {author} {\bibfnamefont {S.~V.}\ \bibnamefont {Morozov}}, \bibinfo
  {author} {\bibfnamefont {D.}~\bibnamefont {Jiang}}, \bibinfo {author}
  {\bibfnamefont {Y.}~\bibnamefont {Zhang}}, \bibinfo {author} {\bibfnamefont
  {S.~V.}\ \bibnamefont {Dubonos}}, \bibinfo {author} {\bibfnamefont {I.~V.}\
  \bibnamefont {Grigorieva}},\ and\ \bibinfo {author} {\bibfnamefont {A.~A.}\
  \bibnamefont {Firsov}},\ }\bibfield  {title} {\bibinfo {title} {Electric
  field effect in atomically thin carbon films},\ }\href
  {https://doi.org/10.1126/science.1102896} {\bibfield  {journal} {\bibinfo
  {journal} {Science}\ }\textbf {\bibinfo {volume} {306}},\ \bibinfo {pages}
  {666} (\bibinfo {year} {2004})}\BibitemShut {NoStop}%
\bibitem [{\citenamefont {Novoselov}\ \emph {et~al.}(2005)\citenamefont
  {Novoselov}, \citenamefont {Geim}, \citenamefont {Morozov}, \citenamefont
  {Jiang}, \citenamefont {Katsnelson}, \citenamefont {Grigorieva},
  \citenamefont {Dubonos},\ and\ \citenamefont {Firsov}}]{novoselov2005two}%
  \BibitemOpen
  \bibfield  {author} {\bibinfo {author} {\bibfnamefont {K.}~\bibnamefont
  {Novoselov}}, \bibinfo {author} {\bibfnamefont {A.}~\bibnamefont {Geim}},
  \bibinfo {author} {\bibfnamefont {S.}~\bibnamefont {Morozov}}, \bibinfo
  {author} {\bibfnamefont {D.}~\bibnamefont {Jiang}}, \bibinfo {author}
  {\bibfnamefont {M.}~\bibnamefont {Katsnelson}}, \bibinfo {author}
  {\bibfnamefont {I.}~\bibnamefont {Grigorieva}}, \bibinfo {author}
  {\bibfnamefont {S.}~\bibnamefont {Dubonos}},\ and\ \bibinfo {author}
  {\bibfnamefont {A.}~\bibnamefont {Firsov}},\ }\bibfield  {title} {\bibinfo
  {title} {Two-dimensional gas of massless dirac fermions in graphene},\ }\href
  {https://doi.org/10.1038/nature04233} {\bibfield  {journal} {\bibinfo
  {journal} {nature}\ }\textbf {\bibinfo {volume} {438}},\ \bibinfo {pages}
  {197} (\bibinfo {year} {2005})}\BibitemShut {NoStop}%
\bibitem [{\citenamefont {Zhang}\ \emph {et~al.}(2005)\citenamefont {Zhang},
  \citenamefont {Tan}, \citenamefont {Stormer},\ and\ \citenamefont
  {Kim}}]{zhang2005experimental}%
  \BibitemOpen
  \bibfield  {author} {\bibinfo {author} {\bibfnamefont {Y.}~\bibnamefont
  {Zhang}}, \bibinfo {author} {\bibfnamefont {Y.-W.}\ \bibnamefont {Tan}},
  \bibinfo {author} {\bibfnamefont {H.~L.}\ \bibnamefont {Stormer}},\ and\
  \bibinfo {author} {\bibfnamefont {P.}~\bibnamefont {Kim}},\ }\bibfield
  {title} {\bibinfo {title} {Experimental observation of the quantum hall
  effect and berry's phase in graphene},\ }\href
  {https://doi.org/10.1038/nature04235} {\bibfield  {journal} {\bibinfo
  {journal} {nature}\ }\textbf {\bibinfo {volume} {438}},\ \bibinfo {pages}
  {201} (\bibinfo {year} {2005})}\BibitemShut {NoStop}%
\bibitem [{\citenamefont {Radisavljevic}\ \emph {et~al.}(2011)\citenamefont
  {Radisavljevic}, \citenamefont {Radenovic}, \citenamefont {Brivio},
  \citenamefont {Giacometti},\ and\ \citenamefont
  {Kis}}]{radisavljevic2011single}%
  \BibitemOpen
  \bibfield  {author} {\bibinfo {author} {\bibfnamefont {B.}~\bibnamefont
  {Radisavljevic}}, \bibinfo {author} {\bibfnamefont {A.}~\bibnamefont
  {Radenovic}}, \bibinfo {author} {\bibfnamefont {J.}~\bibnamefont {Brivio}},
  \bibinfo {author} {\bibfnamefont {V.}~\bibnamefont {Giacometti}},\ and\
  \bibinfo {author} {\bibfnamefont {A.}~\bibnamefont {Kis}},\ }\bibfield
  {title} {\bibinfo {title} {Single-layer mos2 transistors},\ }\href
  {https://doi.org/10.1038/nnano.2010.279} {\bibfield  {journal} {\bibinfo
  {journal} {Nature nanotechnology}\ }\textbf {\bibinfo {volume} {6}},\
  \bibinfo {pages} {147} (\bibinfo {year} {2011})}\BibitemShut {NoStop}%
\bibitem [{\citenamefont {Liu}\ \emph {et~al.}(2017)\citenamefont {Liu},
  \citenamefont {Liu}, \citenamefont {She}, \citenamefont {Zha}, \citenamefont
  {Pan}, \citenamefont {Li}, \citenamefont {Li}, \citenamefont {He},
  \citenamefont {Cai}, \citenamefont {Wang}, \citenamefont {Zheng},
  \citenamefont {Qiu},\ and\ \citenamefont {Zhong}}]{10.1038/ncomms14924}%
  \BibitemOpen
  \bibfield  {author} {\bibinfo {author} {\bibfnamefont {M.}~\bibnamefont
  {Liu}}, \bibinfo {author} {\bibfnamefont {M.}~\bibnamefont {Liu}}, \bibinfo
  {author} {\bibfnamefont {L.}~\bibnamefont {She}}, \bibinfo {author}
  {\bibfnamefont {Z.}~\bibnamefont {Zha}}, \bibinfo {author} {\bibfnamefont
  {J.}~\bibnamefont {Pan}}, \bibinfo {author} {\bibfnamefont {S.}~\bibnamefont
  {Li}}, \bibinfo {author} {\bibfnamefont {T.}~\bibnamefont {Li}}, \bibinfo
  {author} {\bibfnamefont {Y.}~\bibnamefont {He}}, \bibinfo {author}
  {\bibfnamefont {Z.}~\bibnamefont {Cai}}, \bibinfo {author} {\bibfnamefont
  {J.}~\bibnamefont {Wang}}, \bibinfo {author} {\bibfnamefont {Y.}~\bibnamefont
  {Zheng}}, \bibinfo {author} {\bibfnamefont {X.}~\bibnamefont {Qiu}},\ and\
  \bibinfo {author} {\bibfnamefont {D.}~\bibnamefont {Zhong}},\ }\bibfield
  {title} {\bibinfo {title} {Graphene-like nanoribbons periodically embedded
  with four- and eight-membered rings},\ }\href
  {https://doi.org/10.1038/ncomms14924} {\bibfield  {journal} {\bibinfo
  {journal} {Nature Communications}\ }\textbf {\bibinfo {volume} {8}},\
  \bibinfo {pages} {14924} (\bibinfo {year} {2017})}\BibitemShut {NoStop}%
\bibitem [{\citenamefont {Fan}\ \emph {et~al.}(2021)\citenamefont {Fan},
  \citenamefont {Yan}, \citenamefont {Tripp}, \citenamefont {Krejčí},
  \citenamefont {Dimosthenous}, \citenamefont {Kachel}, \citenamefont {Chen},
  \citenamefont {Foster}, \citenamefont {Koert}, \citenamefont {Liljeroth},\
  and\ \citenamefont {Gottfried}}]{10.1126/science.abg4509}%
  \BibitemOpen
  \bibfield  {author} {\bibinfo {author} {\bibfnamefont {Q.}~\bibnamefont
  {Fan}}, \bibinfo {author} {\bibfnamefont {L.}~\bibnamefont {Yan}}, \bibinfo
  {author} {\bibfnamefont {M.~W.}\ \bibnamefont {Tripp}}, \bibinfo {author}
  {\bibfnamefont {O.}~\bibnamefont {Krejčí}}, \bibinfo {author}
  {\bibfnamefont {S.}~\bibnamefont {Dimosthenous}}, \bibinfo {author}
  {\bibfnamefont {S.~R.}\ \bibnamefont {Kachel}}, \bibinfo {author}
  {\bibfnamefont {M.}~\bibnamefont {Chen}}, \bibinfo {author} {\bibfnamefont
  {A.~S.}\ \bibnamefont {Foster}}, \bibinfo {author} {\bibfnamefont
  {U.}~\bibnamefont {Koert}}, \bibinfo {author} {\bibfnamefont
  {P.}~\bibnamefont {Liljeroth}},\ and\ \bibinfo {author} {\bibfnamefont
  {J.~M.}\ \bibnamefont {Gottfried}},\ }\bibfield  {title} {\bibinfo {title}
  {Biphenylene network: A nonbenzenoid carbon allotrope},\ }\href
  {https://doi.org/10.1126/science.abg4509} {\bibfield  {journal} {\bibinfo
  {journal} {Science}\ }\textbf {\bibinfo {volume} {372}},\ \bibinfo {pages}
  {852} (\bibinfo {year} {2021})}\BibitemShut {NoStop}%
\bibitem [{\citenamefont {Hudspeth}\ \emph {et~al.}(2010)\citenamefont
  {Hudspeth}, \citenamefont {Whitman}, \citenamefont {Barone},\ and\
  \citenamefont {Peralta}}]{doi:10.1021/nn100758h}%
  \BibitemOpen
  \bibfield  {author} {\bibinfo {author} {\bibfnamefont {M.~A.}\ \bibnamefont
  {Hudspeth}}, \bibinfo {author} {\bibfnamefont {B.~W.}\ \bibnamefont
  {Whitman}}, \bibinfo {author} {\bibfnamefont {V.}~\bibnamefont {Barone}},\
  and\ \bibinfo {author} {\bibfnamefont {J.~E.}\ \bibnamefont {Peralta}},\
  }\bibfield  {title} {\bibinfo {title} {Electronic properties of the
  biphenylene sheet and its one-dimensional derivatives},\ }\href
  {https://doi.org/10.1021/nn100758h} {\bibfield  {journal} {\bibinfo
  {journal} {ACS Nano}\ }\textbf {\bibinfo {volume} {4}},\ \bibinfo {pages}
  {4565} (\bibinfo {year} {2010})}\BibitemShut {NoStop}%
\bibitem [{\citenamefont {Ge}\ \emph {et~al.}(2016)\citenamefont {Ge},
  \citenamefont {Wang},\ and\ \citenamefont {Liao}}]{GE201697}%
  \BibitemOpen
  \bibfield  {author} {\bibinfo {author} {\bibfnamefont {H.}~\bibnamefont
  {Ge}}, \bibinfo {author} {\bibfnamefont {G.}~\bibnamefont {Wang}},\ and\
  \bibinfo {author} {\bibfnamefont {Y.}~\bibnamefont {Liao}},\ }\bibfield
  {title} {\bibinfo {title} {A theoretical investigation on the transport
  properties of armchair biphenylene nanoribbons},\ }\href
  {https://doi.org/https://doi.org/10.1016/j.cplett.2016.02.011} {\bibfield
  {journal} {\bibinfo  {journal} {Chemical Physics Letters}\ }\textbf {\bibinfo
  {volume} {648}},\ \bibinfo {pages} {97} (\bibinfo {year} {2016})}\BibitemShut
  {NoStop}%
\bibitem [{\citenamefont {Ye}\ \emph {et~al.}(2023)\citenamefont {Ye},
  \citenamefont {Li}, \citenamefont {Zhong},\ and\ \citenamefont
  {Yao}}]{Ye_2023}%
  \BibitemOpen
  \bibfield  {author} {\bibinfo {author} {\bibfnamefont {J.}~\bibnamefont
  {Ye}}, \bibinfo {author} {\bibfnamefont {J.}~\bibnamefont {Li}}, \bibinfo
  {author} {\bibfnamefont {D.}~\bibnamefont {Zhong}},\ and\ \bibinfo {author}
  {\bibfnamefont {D.-X.}\ \bibnamefont {Yao}},\ }\bibfield  {title} {\bibinfo
  {title} {Possible superconductivity in biphenylene},\ }\href
  {https://doi.org/10.1088/0256-307X/40/7/077401} {\bibfield  {journal}
  {\bibinfo  {journal} {Chinese Physics Letters}\ }\textbf {\bibinfo {volume}
  {40}},\ \bibinfo {pages} {077401} (\bibinfo {year} {2023})}\BibitemShut
  {NoStop}%
\bibitem [{\citenamefont {Li}\ and\ \citenamefont {Yao}(2022)}]{JunLi17403}%
  \BibitemOpen
  \bibfield  {author} {\bibinfo {author} {\bibfnamefont {J.}~\bibnamefont
  {Li}}\ and\ \bibinfo {author} {\bibfnamefont {D.-X.}\ \bibnamefont {Yao}},\
  }\bibfield  {title} {\bibinfo {title} {Superconductivity in octagraphene},\
  }\href {https://doi.org/10.1088/1674-1056/ac40fa} {\bibfield  {journal}
  {\bibinfo  {journal} {Chinese Physics B}\ }\textbf {\bibinfo {volume} {31}},\
  \bibinfo {eid} {017403} (\bibinfo {year} {2022})}\BibitemShut {NoStop}%
\bibitem [{\citenamefont {Li}\ \emph {et~al.}(2020)\citenamefont {Li},
  \citenamefont {Jin}, \citenamefont {Yang},\ and\ \citenamefont
  {Yao}}]{PhysRevB.102.174509}%
  \BibitemOpen
  \bibfield  {author} {\bibinfo {author} {\bibfnamefont {J.}~\bibnamefont
  {Li}}, \bibinfo {author} {\bibfnamefont {S.}~\bibnamefont {Jin}}, \bibinfo
  {author} {\bibfnamefont {F.}~\bibnamefont {Yang}},\ and\ \bibinfo {author}
  {\bibfnamefont {D.-X.}\ \bibnamefont {Yao}},\ }\bibfield  {title} {\bibinfo
  {title} {Electronic structure, magnetism, and high-temperature
  superconductivity in multilayer octagraphene and octagraphite},\ }\href
  {https://doi.org/10.1103/PhysRevB.102.174509} {\bibfield  {journal} {\bibinfo
   {journal} {Phys. Rev. B}\ }\textbf {\bibinfo {volume} {102}},\ \bibinfo
  {pages} {174509} (\bibinfo {year} {2020})}\BibitemShut {NoStop}%
\bibitem [{\citenamefont {Kang}\ \emph {et~al.}(2019)\citenamefont {Kang},
  \citenamefont {Lu}, \citenamefont {Yang},\ and\ \citenamefont
  {Yao}}]{PhysRevB.99.184506}%
  \BibitemOpen
  \bibfield  {author} {\bibinfo {author} {\bibfnamefont {Y.-T.}\ \bibnamefont
  {Kang}}, \bibinfo {author} {\bibfnamefont {C.}~\bibnamefont {Lu}}, \bibinfo
  {author} {\bibfnamefont {F.}~\bibnamefont {Yang}},\ and\ \bibinfo {author}
  {\bibfnamefont {D.-X.}\ \bibnamefont {Yao}},\ }\bibfield  {title} {\bibinfo
  {title} {Single-orbital realization of high-temperature
  ${s}^{\ifmmode\pm\else\textpm\fi{}}$ superconductivity in the square-octagon
  lattice},\ }\href {https://doi.org/10.1103/PhysRevB.99.184506} {\bibfield
  {journal} {\bibinfo  {journal} {Phys. Rev. B}\ }\textbf {\bibinfo {volume}
  {99}},\ \bibinfo {pages} {184506} (\bibinfo {year} {2019})}\BibitemShut
  {NoStop}%
\bibitem [{\citenamefont {Zhang}\ and\ \citenamefont
  {Yao}(2023)}]{PhysRevB.108.144407}%
  \BibitemOpen
  \bibfield  {author} {\bibinfo {author} {\bibfnamefont {M.-H.}\ \bibnamefont
  {Zhang}}\ and\ \bibinfo {author} {\bibfnamefont {D.-X.}\ \bibnamefont
  {Yao}},\ }\bibfield  {title} {\bibinfo {title} {Type-{II} dirac points and
  dirac nodal loops on the magnons of a square-hexagon-octagon lattice},\
  }\href {https://doi.org/10.1103/PhysRevB.108.144407} {\bibfield  {journal}
  {\bibinfo  {journal} {Phys. Rev. B}\ }\textbf {\bibinfo {volume} {108}},\
  \bibinfo {pages} {144407} (\bibinfo {year} {2023})}\BibitemShut {NoStop}%
\bibitem [{\citenamefont {Zhao}\ and\ \citenamefont
  {Paramekanti}(2006)}]{PhysRevLett.97.230404}%
  \BibitemOpen
  \bibfield  {author} {\bibinfo {author} {\bibfnamefont {E.}~\bibnamefont
  {Zhao}}\ and\ \bibinfo {author} {\bibfnamefont {A.}~\bibnamefont
  {Paramekanti}},\ }\bibfield  {title} {\bibinfo {title} {Bcs-bec crossover on
  the two-dimensional honeycomb lattice},\ }\href
  {https://doi.org/10.1103/PhysRevLett.97.230404} {\bibfield  {journal}
  {\bibinfo  {journal} {Phys. Rev. Lett.}\ }\textbf {\bibinfo {volume} {97}},\
  \bibinfo {pages} {230404} (\bibinfo {year} {2006})}\BibitemShut {NoStop}%
\bibitem [{\citenamefont {Cichy}\ \emph {et~al.}(2022)\citenamefont {Cichy},
  \citenamefont {Kapcia},\ and\ \citenamefont {Ptok}}]{PhysRevB.105.214510}%
  \BibitemOpen
  \bibfield  {author} {\bibinfo {author} {\bibfnamefont {A.}~\bibnamefont
  {Cichy}}, \bibinfo {author} {\bibfnamefont {K.~J.}\ \bibnamefont {Kapcia}},\
  and\ \bibinfo {author} {\bibfnamefont {A.}~\bibnamefont {Ptok}},\ }\bibfield
  {title} {\bibinfo {title} {Connection between the
  semiconductor-superconductor transition and the spin-polarized
  superconducting phase in the honeycomb lattice},\ }\href
  {https://doi.org/10.1103/PhysRevB.105.214510} {\bibfield  {journal} {\bibinfo
   {journal} {Phys. Rev. B}\ }\textbf {\bibinfo {volume} {105}},\ \bibinfo
  {pages} {214510} (\bibinfo {year} {2022})}\BibitemShut {NoStop}%
\bibitem [{\citenamefont {Cichy}\ \emph {et~al.}(2024)\citenamefont {Cichy},
  \citenamefont {Kapcia},\ and\ \citenamefont {Ptok}}]{CICHY2024171522}%
  \BibitemOpen
  \bibfield  {author} {\bibinfo {author} {\bibfnamefont {A.}~\bibnamefont
  {Cichy}}, \bibinfo {author} {\bibfnamefont {K.~J.}\ \bibnamefont {Kapcia}},\
  and\ \bibinfo {author} {\bibfnamefont {A.}~\bibnamefont {Ptok}},\ }\bibfield
  {title} {\bibinfo {title} {Spin-polarized superconducting phase in
  semiconducting system with next-nearest-neighbor hopping on the honeycomb
  lattice},\ }\href
  {https://doi.org/https://doi.org/10.1016/j.jmmm.2023.171522} {\bibfield
  {journal} {\bibinfo  {journal} {Journal of Magnetism and Magnetic Materials}\
  }\textbf {\bibinfo {volume} {589}},\ \bibinfo {pages} {171522} (\bibinfo
  {year} {2024})}\BibitemShut {NoStop}%
\bibitem [{\citenamefont {Meng}\ \emph {et~al.}(2010)\citenamefont {Meng},
  \citenamefont {Lang}, \citenamefont {Wessel}, \citenamefont {Assaad},\ and\
  \citenamefont {Muramatsu}}]{meng2010quantum}%
  \BibitemOpen
  \bibfield  {author} {\bibinfo {author} {\bibfnamefont {Z.}~\bibnamefont
  {Meng}}, \bibinfo {author} {\bibfnamefont {T.}~\bibnamefont {Lang}}, \bibinfo
  {author} {\bibfnamefont {S.}~\bibnamefont {Wessel}}, \bibinfo {author}
  {\bibfnamefont {F.}~\bibnamefont {Assaad}},\ and\ \bibinfo {author}
  {\bibfnamefont {A.}~\bibnamefont {Muramatsu}},\ }\bibfield  {title} {\bibinfo
  {title} {Quantum spin liquid emerging in two-dimensional correlated dirac
  fermions},\ }\href {https://doi.org/https://doi.org/10.1038/nature08942}
  {\bibfield  {journal} {\bibinfo  {journal} {Nature}\ }\textbf {\bibinfo
  {volume} {464}},\ \bibinfo {pages} {847} (\bibinfo {year}
  {2010})}\BibitemShut {NoStop}%
\bibitem [{\citenamefont {Sorella}\ \emph {et~al.}(2012)\citenamefont
  {Sorella}, \citenamefont {Otsuka},\ and\ \citenamefont
  {Yunoki}}]{sorella2012absence}%
  \BibitemOpen
  \bibfield  {author} {\bibinfo {author} {\bibfnamefont {S.}~\bibnamefont
  {Sorella}}, \bibinfo {author} {\bibfnamefont {Y.}~\bibnamefont {Otsuka}},\
  and\ \bibinfo {author} {\bibfnamefont {S.}~\bibnamefont {Yunoki}},\
  }\bibfield  {title} {\bibinfo {title} {Absence of a spin liquid phase in the
  hubbard model on the honeycomb lattice},\ }\href
  {https://doi.org/10.1038/srep00992} {\bibfield  {journal} {\bibinfo
  {journal} {SCIENTIFIC REPORTS}\ }\textbf {\bibinfo {volume} {2}},\ \bibinfo
  {pages} {992} (\bibinfo {year} {2012})}\BibitemShut {NoStop}%
\bibitem [{\citenamefont {Otsuka}\ \emph {et~al.}(2022)\citenamefont {Otsuka},
  \citenamefont {Seki}, \citenamefont {Sorella},\ and\ \citenamefont
  {Yunoki}}]{Otsuka_2022}%
  \BibitemOpen
  \bibfield  {author} {\bibinfo {author} {\bibfnamefont {Y.}~\bibnamefont
  {Otsuka}}, \bibinfo {author} {\bibfnamefont {K.}~\bibnamefont {Seki}},
  \bibinfo {author} {\bibfnamefont {S.}~\bibnamefont {Sorella}},\ and\ \bibinfo
  {author} {\bibfnamefont {S.}~\bibnamefont {Yunoki}},\ }\bibfield  {title}
  {\bibinfo {title} {{QMC study of the chiral Heisenberg Gross-Neveu
  universality class}},\ }\href
  {https://doi.org/10.1088/1742-6596/2207/1/012030} {\bibfield  {journal}
  {\bibinfo  {journal} {Journal of Physics: Conference Series}\ }\textbf
  {\bibinfo {volume} {2207}},\ \bibinfo {pages} {012030} (\bibinfo {year}
  {2022})}\BibitemShut {NoStop}%
\bibitem [{\citenamefont {Wu}\ and\ \citenamefont
  {Zhang}(2005)}]{PhysRevB.71.155115}%
  \BibitemOpen
  \bibfield  {author} {\bibinfo {author} {\bibfnamefont {C.}~\bibnamefont
  {Wu}}\ and\ \bibinfo {author} {\bibfnamefont {S.-C.}\ \bibnamefont {Zhang}},\
  }\bibfield  {title} {\bibinfo {title} {Sufficient condition for absence of
  the sign problem in the fermionic quantum monte carlo algorithm},\ }\href
  {https://doi.org/10.1103/PhysRevB.71.155115} {\bibfield  {journal} {\bibinfo
  {journal} {Phys. Rev. B}\ }\textbf {\bibinfo {volume} {71}},\ \bibinfo
  {pages} {155115} (\bibinfo {year} {2005})}\BibitemShut {NoStop}%
\bibitem [{\citenamefont {Assaad}\ and\ \citenamefont
  {Evertz}(2008)}]{Assaad2008}%
  \BibitemOpen
  \bibfield  {author} {\bibinfo {author} {\bibfnamefont {F.~F.}\ \bibnamefont
  {Assaad}}\ and\ \bibinfo {author} {\bibfnamefont {H.}~\bibnamefont
  {Evertz}},\ }\bibfield  {title} {\bibinfo {title} {World-line and
  determinantal quantum monte carlo methods for spins, phonons and electrons},\
  }in\ \href {https://doi.org/10.1007/978-3-540-74686-7_10} {\emph {\bibinfo
  {booktitle} {Computational Many-Particle Physics}}},\ \bibinfo {editor}
  {edited by\ \bibinfo {editor} {\bibfnamefont {H.}~\bibnamefont {Fehske}},
  \bibinfo {editor} {\bibfnamefont {R.}~\bibnamefont {Schneider}},\ and\
  \bibinfo {editor} {\bibfnamefont {A.}~\bibnamefont {Wei{\ss}e}}}\ (\bibinfo
  {publisher} {Springer},\ \bibinfo {address} {Berlin, Heidelberg},\ \bibinfo
  {year} {2008})\ pp.\ \bibinfo {pages} {277--356}\BibitemShut {NoStop}%
\bibitem [{\citenamefont {Otsuka}\ \emph {et~al.}(2016)\citenamefont {Otsuka},
  \citenamefont {Yunoki},\ and\ \citenamefont {Sorella}}]{PhysRevX.6.011029}%
  \BibitemOpen
  \bibfield  {author} {\bibinfo {author} {\bibfnamefont {Y.}~\bibnamefont
  {Otsuka}}, \bibinfo {author} {\bibfnamefont {S.}~\bibnamefont {Yunoki}},\
  and\ \bibinfo {author} {\bibfnamefont {S.}~\bibnamefont {Sorella}},\
  }\bibfield  {title} {\bibinfo {title} {Universal quantum criticality in the
  metal-insulator transition of two-dimensional interacting dirac electrons},\
  }\href {https://doi.org/10.1103/PhysRevX.6.011029} {\bibfield  {journal}
  {\bibinfo  {journal} {Phys. Rev. X}\ }\textbf {\bibinfo {volume} {6}},\
  \bibinfo {pages} {011029} (\bibinfo {year} {2016})}\BibitemShut {NoStop}%
\bibitem [{\citenamefont {Sorella}\ and\ \citenamefont
  {Tosatti}(1992)}]{S.Sorella_1992}%
  \BibitemOpen
  \bibfield  {author} {\bibinfo {author} {\bibfnamefont {S.}~\bibnamefont
  {Sorella}}\ and\ \bibinfo {author} {\bibfnamefont {E.}~\bibnamefont
  {Tosatti}},\ }\bibfield  {title} {\bibinfo {title} {Semi-metal-insulator
  transition of the hubbard model in the honeycomb lattice},\ }\href
  {https://doi.org/10.1209/0295-5075/19/8/007} {\bibfield  {journal} {\bibinfo
  {journal} {Europhysics Letters}\ }\textbf {\bibinfo {volume} {19}},\ \bibinfo
  {pages} {699} (\bibinfo {year} {1992})}\BibitemShut {NoStop}%
\bibitem [{\citenamefont {Herbut}(2006)}]{PhysRevLett.97.146401}%
  \BibitemOpen
  \bibfield  {author} {\bibinfo {author} {\bibfnamefont {I.~F.}\ \bibnamefont
  {Herbut}},\ }\bibfield  {title} {\bibinfo {title} {Interactions and phase
  transitions on graphene's honeycomb lattice},\ }\href
  {https://doi.org/10.1103/PhysRevLett.97.146401} {\bibfield  {journal}
  {\bibinfo  {journal} {Phys. Rev. Lett.}\ }\textbf {\bibinfo {volume} {97}},\
  \bibinfo {pages} {146401} (\bibinfo {year} {2006})}\BibitemShut {NoStop}%
\bibitem [{\citenamefont {Herbut}\ \emph {et~al.}(2009)\citenamefont {Herbut},
  \citenamefont {Juri\ifmmode \check{c}\else \v{c}\fi{}i\ifmmode~\acute{c}\else
  \'{c}\fi{}},\ and\ \citenamefont {Vafek}}]{PhysRevB.80.075432}%
  \BibitemOpen
  \bibfield  {author} {\bibinfo {author} {\bibfnamefont {I.~F.}\ \bibnamefont
  {Herbut}}, \bibinfo {author} {\bibfnamefont {V.}~\bibnamefont {Juri\ifmmode
  \check{c}\else \v{c}\fi{}i\ifmmode~\acute{c}\else \'{c}\fi{}}},\ and\
  \bibinfo {author} {\bibfnamefont {O.}~\bibnamefont {Vafek}},\ }\bibfield
  {title} {\bibinfo {title} {Relativistic mott criticality in graphene},\
  }\href {https://doi.org/10.1103/PhysRevB.80.075432} {\bibfield  {journal}
  {\bibinfo  {journal} {Phys. Rev. B}\ }\textbf {\bibinfo {volume} {80}},\
  \bibinfo {pages} {075432} (\bibinfo {year} {2009})}\BibitemShut {NoStop}%
\bibitem [{\citenamefont {Assaad}\ and\ \citenamefont
  {Herbut}(2013)}]{PhysRevX.3.031010}%
  \BibitemOpen
  \bibfield  {author} {\bibinfo {author} {\bibfnamefont {F.~F.}\ \bibnamefont
  {Assaad}}\ and\ \bibinfo {author} {\bibfnamefont {I.~F.}\ \bibnamefont
  {Herbut}},\ }\bibfield  {title} {\bibinfo {title} {Pinning the order: The
  nature of quantum criticality in the hubbard model on honeycomb lattice},\
  }\href {https://doi.org/10.1103/PhysRevX.3.031010} {\bibfield  {journal}
  {\bibinfo  {journal} {Phys. Rev. X}\ }\textbf {\bibinfo {volume} {3}},\
  \bibinfo {pages} {031010} (\bibinfo {year} {2013})}\BibitemShut {NoStop}%
\bibitem [{\citenamefont {Yu}\ \emph {et~al.}()\citenamefont {Yu},
  \citenamefont {Zeng}, \citenamefont {Shu}, \citenamefont {Li},\ and\
  \citenamefont {Yin}}]{arxiv231010601}%
  \BibitemOpen
  \bibfield  {author} {\bibinfo {author} {\bibfnamefont {Y.-K.}\ \bibnamefont
  {Yu}}, \bibinfo {author} {\bibfnamefont {Z.}~\bibnamefont {Zeng}}, \bibinfo
  {author} {\bibfnamefont {Y.-R.}\ \bibnamefont {Shu}}, \bibinfo {author}
  {\bibfnamefont {Z.-X.}\ \bibnamefont {Li}},\ and\ \bibinfo {author}
  {\bibfnamefont {S.}~\bibnamefont {Yin}},\ }\bibfield  {title} {\bibinfo
  {title} {Nonequilibrium dynamics in dirac quantum criticality},\ }\href
  {https://doi.org/10.48550/arXiv.2310.10601} {\bibinfo  {journal}
  {arXiv:2310.10601}\ }\BibitemShut {NoStop}%
\bibitem [{\citenamefont {Assaad}\ \emph {et~al.}(2022)\citenamefont {Assaad},
  \citenamefont {Bercx}, \citenamefont {Goth}, \citenamefont {Götz},
  \citenamefont {Hofmann}, \citenamefont {Huffman}, \citenamefont {Liu},
  \citenamefont {Toldin}, \citenamefont {Portela},\ and\ \citenamefont
  {Schwab}}]{Assaad_2022ALF}%
  \BibitemOpen
\bibfield  {journal} {  }\bibfield  {author} {\bibinfo {author} {\bibfnamefont
  {F.}~\bibnamefont {Assaad}}, \bibinfo {author} {\bibfnamefont
  {M.}~\bibnamefont {Bercx}}, \bibinfo {author} {\bibfnamefont
  {F.}~\bibnamefont {Goth}}, \bibinfo {author} {\bibfnamefont {A.}~\bibnamefont
  {Götz}}, \bibinfo {author} {\bibfnamefont {J.}~\bibnamefont {Hofmann}},
  \bibinfo {author} {\bibfnamefont {E.}~\bibnamefont {Huffman}}, \bibinfo
  {author} {\bibfnamefont {Z.}~\bibnamefont {Liu}}, \bibinfo {author}
  {\bibfnamefont {F.~P.}\ \bibnamefont {Toldin}}, \bibinfo {author}
  {\bibfnamefont {J.}~\bibnamefont {Portela}},\ and\ \bibinfo {author}
  {\bibfnamefont {J.}~\bibnamefont {Schwab}},\ }\bibfield  {title} {\bibinfo
  {title} {The {ALF} (algorithms for lattice fermions) project release 2.0.
  documentation for the auxiliary-field quantum monte carlo code},\ }\bibfield
  {journal} {\bibinfo  {journal} {{SciPost} Physics Codebases}\ }\href
  {https://doi.org/10.21468/scipostphyscodeb.1} {10.21468/scipostphyscodeb.1}
  (\bibinfo {year} {2022})\BibitemShut {NoStop}%
\bibitem [{\citenamefont {Trotter}(1959)}]{trotter1959product}%
  \BibitemOpen
  \bibfield  {author} {\bibinfo {author} {\bibfnamefont {H.~F.}\ \bibnamefont
  {Trotter}},\ }\bibfield  {title} {\bibinfo {title} {On the product of
  semi-groups of operators},\ }\href@noop {} {\bibfield  {journal} {\bibinfo
  {journal} {Proceedings of the American Mathematical Society}\ }\textbf
  {\bibinfo {volume} {10}},\ \bibinfo {pages} {545} (\bibinfo {year}
  {1959})}\BibitemShut {NoStop}%
\bibitem [{\citenamefont {Suzuki}(1991)}]{suzuki1991general}%
  \BibitemOpen
  \bibfield  {author} {\bibinfo {author} {\bibfnamefont {M.}~\bibnamefont
  {Suzuki}},\ }\bibfield  {title} {\bibinfo {title} {General theory of fractal
  path integrals with applications to many-body theories and statistical
  physics},\ }\href@noop {} {\bibfield  {journal} {\bibinfo  {journal} {Journal
  of Mathematical Physics}\ }\textbf {\bibinfo {volume} {32}},\ \bibinfo
  {pages} {400} (\bibinfo {year} {1991})}\BibitemShut {NoStop}%
\bibitem [{\citenamefont {Hubbard}(1959)}]{PhysRevLett.3.77}%
  \BibitemOpen
  \bibfield  {author} {\bibinfo {author} {\bibfnamefont {J.}~\bibnamefont
  {Hubbard}},\ }\bibfield  {title} {\bibinfo {title} {Calculation of partition
  functions},\ }\href {https://doi.org/10.1103/PhysRevLett.3.77} {\bibfield
  {journal} {\bibinfo  {journal} {Phys. Rev. Lett.}\ }\textbf {\bibinfo
  {volume} {3}},\ \bibinfo {pages} {77} (\bibinfo {year} {1959})}\BibitemShut
  {NoStop}%
\bibitem [{\citenamefont {Sandvik}(1998)}]{PhysRevB.57.10287}%
  \BibitemOpen
  \bibfield  {author} {\bibinfo {author} {\bibfnamefont {A.~W.}\ \bibnamefont
  {Sandvik}},\ }\bibfield  {title} {\bibinfo {title} {Stochastic method for
  analytic continuation of quantum monte carlo data},\ }\href
  {https://link.aps.org/doi/10.1103/PhysRevB.57.10287} {\bibfield  {journal}
  {\bibinfo  {journal} {Phys. Rev. B}\ }\textbf {\bibinfo {volume} {57}},\
  \bibinfo {pages} {10287} (\bibinfo {year} {1998})}\BibitemShut {NoStop}%
\bibitem [{\citenamefont {Beach}()}]{arxiv0403055}%
  \BibitemOpen
  \bibfield  {author} {\bibinfo {author} {\bibfnamefont {K.~S.~D.}\
  \bibnamefont {Beach}},\ }\bibfield  {title} {\bibinfo {title} {Identifying
  the maximum entropy method as a special limit of stochastic analytic
  continuation},\ }\href {https://doi.org/10.48550/arXiv.cond-mat/0403055}
  {\bibinfo  {journal} {arXiv:cond-mat/0403055}\ }\BibitemShut {NoStop}%
\bibitem [{\citenamefont {Singh}\ \emph {et~al.}(1988)\citenamefont {Singh},
  \citenamefont {Gelfand},\ and\ \citenamefont {Huse}}]{PhysRevLett.61.2484}%
  \BibitemOpen
\bibfield  {journal} {  }\bibfield  {author} {\bibinfo {author} {\bibfnamefont
  {R.~R.~P.}\ \bibnamefont {Singh}}, \bibinfo {author} {\bibfnamefont {M.~P.}\
  \bibnamefont {Gelfand}},\ and\ \bibinfo {author} {\bibfnamefont {D.~A.}\
  \bibnamefont {Huse}},\ }\bibfield  {title} {\bibinfo {title} {Ground states
  of low-dimensional quantum antiferromagnets},\ }\href
  {https://doi.org/10.1103/PhysRevLett.61.2484} {\bibfield  {journal} {\bibinfo
   {journal} {Phys. Rev. Lett.}\ }\textbf {\bibinfo {volume} {61}},\ \bibinfo
  {pages} {2484} (\bibinfo {year} {1988})}\BibitemShut {NoStop}%
\bibitem [{\citenamefont {Wenzel}\ \emph {et~al.}(2008)\citenamefont {Wenzel},
  \citenamefont {Bogacz},\ and\ \citenamefont
  {Janke}}]{PhysRevLett.101.127202}%
  \BibitemOpen
  \bibfield  {author} {\bibinfo {author} {\bibfnamefont {S.}~\bibnamefont
  {Wenzel}}, \bibinfo {author} {\bibfnamefont {L.}~\bibnamefont {Bogacz}},\
  and\ \bibinfo {author} {\bibfnamefont {W.}~\bibnamefont {Janke}},\ }\bibfield
   {title} {\bibinfo {title} {Evidence for an unconventional universality class
  from a two-dimensional dimerized quantum heisenberg model},\ }\href
  {https://doi.org/10.1103/PhysRevLett.101.127202} {\bibfield  {journal}
  {\bibinfo  {journal} {Phys. Rev. Lett.}\ }\textbf {\bibinfo {volume} {101}},\
  \bibinfo {pages} {127202} (\bibinfo {year} {2008})}\BibitemShut {NoStop}%
\bibitem [{\citenamefont {Troyer}\ \emph {et~al.}(1997)\citenamefont {Troyer},
  \citenamefont {Imada},\ and\ \citenamefont
  {Ueda}}]{doi:10.1143/JPSJ.66.2957}%
  \BibitemOpen
  \bibfield  {author} {\bibinfo {author} {\bibfnamefont {M.}~\bibnamefont
  {Troyer}}, \bibinfo {author} {\bibfnamefont {M.}~\bibnamefont {Imada}},\ and\
  \bibinfo {author} {\bibfnamefont {K.}~\bibnamefont {Ueda}},\ }\bibfield
  {title} {\bibinfo {title} {Critical exponents of the quantum phase transition
  in a planar antiferromagnet},\ }\href {https://doi.org/10.1143/JPSJ.66.2957}
  {\bibfield  {journal} {\bibinfo  {journal} {Journal of the Physical Society
  of Japan}\ }\textbf {\bibinfo {volume} {66}},\ \bibinfo {pages} {2957}
  (\bibinfo {year} {1997})}\BibitemShut {NoStop}%
\bibitem [{\citenamefont {Matsumoto}\ \emph {et~al.}(2001)\citenamefont
  {Matsumoto}, \citenamefont {Yasuda}, \citenamefont {Todo},\ and\
  \citenamefont {Takayama}}]{PhysRevB.65.014407}%
  \BibitemOpen
  \bibfield  {author} {\bibinfo {author} {\bibfnamefont {M.}~\bibnamefont
  {Matsumoto}}, \bibinfo {author} {\bibfnamefont {C.}~\bibnamefont {Yasuda}},
  \bibinfo {author} {\bibfnamefont {S.}~\bibnamefont {Todo}},\ and\ \bibinfo
  {author} {\bibfnamefont {H.}~\bibnamefont {Takayama}},\ }\bibfield  {title}
  {\bibinfo {title} {Ground-state phase diagram of quantum heisenberg
  antiferromagnets on the anisotropic dimerized square lattice},\ }\href
  {https://doi.org/10.1103/PhysRevB.65.014407} {\bibfield  {journal} {\bibinfo
  {journal} {Phys. Rev. B}\ }\textbf {\bibinfo {volume} {65}},\ \bibinfo
  {pages} {014407} (\bibinfo {year} {2001})}\BibitemShut {NoStop}%
\bibitem [{\citenamefont {Wenzel}\ and\ \citenamefont
  {Janke}(2009)}]{PhysRevB.79.014410}%
  \BibitemOpen
  \bibfield  {author} {\bibinfo {author} {\bibfnamefont {S.}~\bibnamefont
  {Wenzel}}\ and\ \bibinfo {author} {\bibfnamefont {W.}~\bibnamefont {Janke}},\
  }\bibfield  {title} {\bibinfo {title} {Comprehensive quantum monte carlo
  study of the quantum critical points in planar dimerized/quadrumerized
  heisenberg models},\ }\href {https://doi.org/10.1103/PhysRevB.79.014410}
  {\bibfield  {journal} {\bibinfo  {journal} {Phys. Rev. B}\ }\textbf {\bibinfo
  {volume} {79}},\ \bibinfo {pages} {014410} (\bibinfo {year}
  {2009})}\BibitemShut {NoStop}%
\bibitem [{\citenamefont {Ran}\ \emph {et~al.}(2019)\citenamefont {Ran},
  \citenamefont {Ma},\ and\ \citenamefont {Yao}}]{PhysRevB.99.174434}%
  \BibitemOpen
  \bibfield  {author} {\bibinfo {author} {\bibfnamefont {X.}~\bibnamefont
  {Ran}}, \bibinfo {author} {\bibfnamefont {N.}~\bibnamefont {Ma}},\ and\
  \bibinfo {author} {\bibfnamefont {D.-X.}\ \bibnamefont {Yao}},\ }\bibfield
  {title} {\bibinfo {title} {Criticality and scaling corrections for
  two-dimensional heisenberg models in plaquette patterns with strong and weak
  couplings},\ }\href {https://doi.org/10.1103/PhysRevB.99.174434} {\bibfield
  {journal} {\bibinfo  {journal} {Phys. Rev. B}\ }\textbf {\bibinfo {volume}
  {99}},\ \bibinfo {pages} {174434} (\bibinfo {year} {2019})}\BibitemShut
  {NoStop}%
\bibitem [{\citenamefont {Wang}\ \emph {et~al.}(2006)\citenamefont {Wang},
  \citenamefont {Beach},\ and\ \citenamefont {Sandvik}}]{PhysRevB.73.014431}%
  \BibitemOpen
  \bibfield  {author} {\bibinfo {author} {\bibfnamefont {L.}~\bibnamefont
  {Wang}}, \bibinfo {author} {\bibfnamefont {K.~S.~D.}\ \bibnamefont {Beach}},\
  and\ \bibinfo {author} {\bibfnamefont {A.~W.}\ \bibnamefont {Sandvik}},\
  }\bibfield  {title} {\bibinfo {title} {High-precision finite-size scaling
  analysis of the quantum-critical point of {S=1/2} heisenberg
  antiferromagnetic bilayers},\ }\href
  {https://doi.org/10.1103/PhysRevB.73.014431} {\bibfield  {journal} {\bibinfo
  {journal} {Phys. Rev. B}\ }\textbf {\bibinfo {volume} {73}},\ \bibinfo
  {pages} {014431} (\bibinfo {year} {2006})}\BibitemShut {NoStop}%
\bibitem [{\citenamefont {Campostrini}\ \emph {et~al.}(2002)\citenamefont
  {Campostrini}, \citenamefont {Hasenbusch}, \citenamefont {Pelissetto},
  \citenamefont {Rossi},\ and\ \citenamefont {Vicari}}]{PhysRevB.65.144520}%
  \BibitemOpen
  \bibfield  {author} {\bibinfo {author} {\bibfnamefont {M.}~\bibnamefont
  {Campostrini}}, \bibinfo {author} {\bibfnamefont {M.}~\bibnamefont
  {Hasenbusch}}, \bibinfo {author} {\bibfnamefont {A.}~\bibnamefont
  {Pelissetto}}, \bibinfo {author} {\bibfnamefont {P.}~\bibnamefont {Rossi}},\
  and\ \bibinfo {author} {\bibfnamefont {E.}~\bibnamefont {Vicari}},\
  }\bibfield  {title} {\bibinfo {title} {Critical exponents and equation of
  state of the three-dimensional heisenberg universality class},\ }\href
  {https://doi.org/10.1103/PhysRevB.65.144520} {\bibfield  {journal} {\bibinfo
  {journal} {Phys. Rev. B}\ }\textbf {\bibinfo {volume} {65}},\ \bibinfo
  {pages} {144520} (\bibinfo {year} {2002})}\BibitemShut {NoStop}%
\bibitem [{\citenamefont {Soluyanov}\ \emph {et~al.}(2015)\citenamefont
  {Soluyanov}, \citenamefont {Gresch}, \citenamefont {Wang}, \citenamefont
  {Wu}, \citenamefont {Troyer}, \citenamefont {Dai},\ and\ \citenamefont
  {Bernevig}}]{soluyanov2015type}%
  \BibitemOpen
  \bibfield  {author} {\bibinfo {author} {\bibfnamefont {A.~A.}\ \bibnamefont
  {Soluyanov}}, \bibinfo {author} {\bibfnamefont {D.}~\bibnamefont {Gresch}},
  \bibinfo {author} {\bibfnamefont {Z.}~\bibnamefont {Wang}}, \bibinfo {author}
  {\bibfnamefont {Q.}~\bibnamefont {Wu}}, \bibinfo {author} {\bibfnamefont
  {M.}~\bibnamefont {Troyer}}, \bibinfo {author} {\bibfnamefont
  {X.}~\bibnamefont {Dai}},\ and\ \bibinfo {author} {\bibfnamefont {B.~A.}\
  \bibnamefont {Bernevig}},\ }\bibfield  {title} {\bibinfo {title} {Type-{II}
  weyl semimetals},\ }\href {https://doi.org/10.1038/nature15768} {\bibfield
  {journal} {\bibinfo  {journal} {Nature}\ }\textbf {\bibinfo {volume} {527}},\
  \bibinfo {pages} {495} (\bibinfo {year} {2015})}\BibitemShut {NoStop}%
\bibitem [{\citenamefont {Xu}\ \emph {et~al.}(2015)\citenamefont {Xu},
  \citenamefont {Zhang},\ and\ \citenamefont {Zhang}}]{PhysRevLett.115.265304}%
  \BibitemOpen
  \bibfield  {author} {\bibinfo {author} {\bibfnamefont {Y.}~\bibnamefont
  {Xu}}, \bibinfo {author} {\bibfnamefont {F.}~\bibnamefont {Zhang}},\ and\
  \bibinfo {author} {\bibfnamefont {C.}~\bibnamefont {Zhang}},\ }\bibfield
  {title} {\bibinfo {title} {Structured weyl points in spin-orbit coupled
  fermionic superfluids},\ }\href
  {https://doi.org/10.1103/PhysRevLett.115.265304} {\bibfield  {journal}
  {\bibinfo  {journal} {Phys. Rev. Lett.}\ }\textbf {\bibinfo {volume} {115}},\
  \bibinfo {pages} {265304} (\bibinfo {year} {2015})}\BibitemShut {NoStop}%
\bibitem [{\citenamefont {Chang}\ \emph {et~al.}(2017)\citenamefont {Chang},
  \citenamefont {Xu}, \citenamefont {Sanchez}, \citenamefont {Tsai},
  \citenamefont {Huang}, \citenamefont {Chang}, \citenamefont {Hsu},
  \citenamefont {Bian}, \citenamefont {Belopolski}, \citenamefont {Yu},
  \citenamefont {Yang}, \citenamefont {Neupert}, \citenamefont {Jeng},
  \citenamefont {Lin},\ and\ \citenamefont {Hasan}}]{PhysRevLett.119.026404}%
  \BibitemOpen
  \bibfield  {author} {\bibinfo {author} {\bibfnamefont {T.-R.}\ \bibnamefont
  {Chang}}, \bibinfo {author} {\bibfnamefont {S.-Y.}\ \bibnamefont {Xu}},
  \bibinfo {author} {\bibfnamefont {D.~S.}\ \bibnamefont {Sanchez}}, \bibinfo
  {author} {\bibfnamefont {W.-F.}\ \bibnamefont {Tsai}}, \bibinfo {author}
  {\bibfnamefont {S.-M.}\ \bibnamefont {Huang}}, \bibinfo {author}
  {\bibfnamefont {G.}~\bibnamefont {Chang}}, \bibinfo {author} {\bibfnamefont
  {C.-H.}\ \bibnamefont {Hsu}}, \bibinfo {author} {\bibfnamefont
  {G.}~\bibnamefont {Bian}}, \bibinfo {author} {\bibfnamefont {I.}~\bibnamefont
  {Belopolski}}, \bibinfo {author} {\bibfnamefont {Z.-M.}\ \bibnamefont {Yu}},
  \bibinfo {author} {\bibfnamefont {S.~A.}\ \bibnamefont {Yang}}, \bibinfo
  {author} {\bibfnamefont {T.}~\bibnamefont {Neupert}}, \bibinfo {author}
  {\bibfnamefont {H.-T.}\ \bibnamefont {Jeng}}, \bibinfo {author}
  {\bibfnamefont {H.}~\bibnamefont {Lin}},\ and\ \bibinfo {author}
  {\bibfnamefont {M.~Z.}\ \bibnamefont {Hasan}},\ }\bibfield  {title} {\bibinfo
  {title} {Type-{II} symmetry-protected topological dirac semimetals},\ }\href
  {https://doi.org/10.1103/PhysRevLett.119.026404} {\bibfield  {journal}
  {\bibinfo  {journal} {Phys. Rev. Lett.}\ }\textbf {\bibinfo {volume} {119}},\
  \bibinfo {pages} {026404} (\bibinfo {year} {2017})}\BibitemShut {NoStop}%
\bibitem [{\citenamefont {Yu}\ \emph {et~al.}(2016)\citenamefont {Yu},
  \citenamefont {Yao},\ and\ \citenamefont {Yang}}]{PhysRevLett.117.077202}%
  \BibitemOpen
  \bibfield  {author} {\bibinfo {author} {\bibfnamefont {Z.-M.}\ \bibnamefont
  {Yu}}, \bibinfo {author} {\bibfnamefont {Y.}~\bibnamefont {Yao}},\ and\
  \bibinfo {author} {\bibfnamefont {S.~A.}\ \bibnamefont {Yang}},\ }\bibfield
  {title} {\bibinfo {title} {Predicted unusual magnetoresponse in type-{II}
  weyl semimetals},\ }\href {https://doi.org/10.1103/PhysRevLett.117.077202}
  {\bibfield  {journal} {\bibinfo  {journal} {Phys. Rev. Lett.}\ }\textbf
  {\bibinfo {volume} {117}},\ \bibinfo {pages} {077202} (\bibinfo {year}
  {2016})}\BibitemShut {NoStop}%
\bibitem [{\citenamefont {Tchoumakov}\ \emph {et~al.}(2016)\citenamefont
  {Tchoumakov}, \citenamefont {Civelli},\ and\ \citenamefont
  {Goerbig}}]{PhysRevLett.117.086402}%
  \BibitemOpen
  \bibfield  {author} {\bibinfo {author} {\bibfnamefont {S.}~\bibnamefont
  {Tchoumakov}}, \bibinfo {author} {\bibfnamefont {M.}~\bibnamefont
  {Civelli}},\ and\ \bibinfo {author} {\bibfnamefont {M.~O.}\ \bibnamefont
  {Goerbig}},\ }\bibfield  {title} {\bibinfo {title} {Magnetic-field-induced
  relativistic properties in type-i and type-{II} weyl semimetals},\ }\href
  {https://doi.org/10.1103/PhysRevLett.117.086402} {\bibfield  {journal}
  {\bibinfo  {journal} {Phys. Rev. Lett.}\ }\textbf {\bibinfo {volume} {117}},\
  \bibinfo {pages} {086402} (\bibinfo {year} {2016})}\BibitemShut {NoStop}%
\bibitem [{\citenamefont {Guan}\ \emph {et~al.}(2017)\citenamefont {Guan},
  \citenamefont {Yu}, \citenamefont {Liu}, \citenamefont {Liu}, \citenamefont
  {Dong}, \citenamefont {Lu}, \citenamefont {Yao},\ and\ \citenamefont
  {Yang}}]{guan2017artificial}%
  \BibitemOpen
  \bibfield  {author} {\bibinfo {author} {\bibfnamefont {S.}~\bibnamefont
  {Guan}}, \bibinfo {author} {\bibfnamefont {Z.-M.}\ \bibnamefont {Yu}},
  \bibinfo {author} {\bibfnamefont {Y.}~\bibnamefont {Liu}}, \bibinfo {author}
  {\bibfnamefont {G.-B.}\ \bibnamefont {Liu}}, \bibinfo {author} {\bibfnamefont
  {L.}~\bibnamefont {Dong}}, \bibinfo {author} {\bibfnamefont {Y.}~\bibnamefont
  {Lu}}, \bibinfo {author} {\bibfnamefont {Y.}~\bibnamefont {Yao}},\ and\
  \bibinfo {author} {\bibfnamefont {S.~A.}\ \bibnamefont {Yang}},\ }\bibfield
  {title} {\bibinfo {title} {Artificial gravity field, astrophysical analogues,
  and topological phase transitions in strained topological semimetals},\
  }\href {https://doi.org/10.1038/s41535-017-0026-7} {\bibfield  {journal}
  {\bibinfo  {journal} {npj Quantum Materials}\ }\textbf {\bibinfo {volume}
  {2}},\ \bibinfo {pages} {23} (\bibinfo {year} {2017})}\BibitemShut {NoStop}%
\bibitem [{\citenamefont {Yang}\ and\ \citenamefont
  {Nagaosa}(2014)}]{Yang2014ClassificationOS}%
  \BibitemOpen
  \bibfield  {author} {\bibinfo {author} {\bibfnamefont {B.-J.}\ \bibnamefont
  {Yang}}\ and\ \bibinfo {author} {\bibfnamefont {N.}~\bibnamefont {Nagaosa}},\
  }\bibfield  {title} {\bibinfo {title} {Classification of stable
  three-dimensional dirac semimetals with nontrivial topology},\ }\href
  {https://doi.org/10.1038/ncomms5898} {\bibfield  {journal} {\bibinfo
  {journal} {Nature Communications}\ }\textbf {\bibinfo {volume} {5}},\
  \bibinfo {pages} {4898} (\bibinfo {year} {2014})}\BibitemShut {NoStop}%
\bibitem [{\citenamefont {Wang}\ \emph {et~al.}(2012)\citenamefont {Wang},
  \citenamefont {Sun}, \citenamefont {Chen}, \citenamefont {Franchini},
  \citenamefont {Xu}, \citenamefont {Weng}, \citenamefont {Dai},\ and\
  \citenamefont {Fang}}]{PhysRevB.85.195320}%
  \BibitemOpen
  \bibfield  {author} {\bibinfo {author} {\bibfnamefont {Z.}~\bibnamefont
  {Wang}}, \bibinfo {author} {\bibfnamefont {Y.}~\bibnamefont {Sun}}, \bibinfo
  {author} {\bibfnamefont {X.-Q.}\ \bibnamefont {Chen}}, \bibinfo {author}
  {\bibfnamefont {C.}~\bibnamefont {Franchini}}, \bibinfo {author}
  {\bibfnamefont {G.}~\bibnamefont {Xu}}, \bibinfo {author} {\bibfnamefont
  {H.}~\bibnamefont {Weng}}, \bibinfo {author} {\bibfnamefont {X.}~\bibnamefont
  {Dai}},\ and\ \bibinfo {author} {\bibfnamefont {Z.}~\bibnamefont {Fang}},\
  }\bibfield  {title} {\bibinfo {title} {Dirac semimetal and topological phase
  transitions in {${A}_{3}$Bi ($A=\text{Na}$, K, Rb)}},\ }\href
  {https://doi.org/10.1103/PhysRevB.85.195320} {\bibfield  {journal} {\bibinfo
  {journal} {Phys. Rev. B}\ }\textbf {\bibinfo {volume} {85}},\ \bibinfo
  {pages} {195320} (\bibinfo {year} {2012})}\BibitemShut {NoStop}%
\bibitem [{\citenamefont {Liu}\ \emph {et~al.}(2014)\citenamefont {Liu},
  \citenamefont {Zhou}, \citenamefont {Zhang}, \citenamefont {Wang},
  \citenamefont {Weng}, \citenamefont {Prabhakaran}, \citenamefont {Mo},
  \citenamefont {Shen}, \citenamefont {Fang}, \citenamefont {Dai},
  \citenamefont {Hussain},\ and\ \citenamefont {Chen}}]{science1245085}%
  \BibitemOpen
  \bibfield  {author} {\bibinfo {author} {\bibfnamefont {Z.~K.}\ \bibnamefont
  {Liu}}, \bibinfo {author} {\bibfnamefont {B.}~\bibnamefont {Zhou}}, \bibinfo
  {author} {\bibfnamefont {Y.}~\bibnamefont {Zhang}}, \bibinfo {author}
  {\bibfnamefont {Z.~J.}\ \bibnamefont {Wang}}, \bibinfo {author}
  {\bibfnamefont {H.~M.}\ \bibnamefont {Weng}}, \bibinfo {author}
  {\bibfnamefont {D.}~\bibnamefont {Prabhakaran}}, \bibinfo {author}
  {\bibfnamefont {S.-K.}\ \bibnamefont {Mo}}, \bibinfo {author} {\bibfnamefont
  {Z.~X.}\ \bibnamefont {Shen}}, \bibinfo {author} {\bibfnamefont
  {Z.}~\bibnamefont {Fang}}, \bibinfo {author} {\bibfnamefont {X.}~\bibnamefont
  {Dai}}, \bibinfo {author} {\bibfnamefont {Z.}~\bibnamefont {Hussain}},\ and\
  \bibinfo {author} {\bibfnamefont {Y.~L.}\ \bibnamefont {Chen}},\ }\bibfield
  {title} {\bibinfo {title} {Discovery of a three-dimensional topological dirac
  semimetal, $\text{Na}_{3}\text{Bi}$},\ }\href
  {https://doi.org/10.1126/science.1245085} {\bibfield  {journal} {\bibinfo
  {journal} {Science}\ }\textbf {\bibinfo {volume} {343}},\ \bibinfo {pages}
  {864} (\bibinfo {year} {2014})}\BibitemShut {NoStop}%
\bibitem [{\citenamefont {Wang}\ \emph {et~al.}(2013)\citenamefont {Wang},
  \citenamefont {Weng}, \citenamefont {Wu}, \citenamefont {Dai},\ and\
  \citenamefont {Fang}}]{PhysRevB88125427}%
  \BibitemOpen
  \bibfield  {author} {\bibinfo {author} {\bibfnamefont {Z.}~\bibnamefont
  {Wang}}, \bibinfo {author} {\bibfnamefont {H.}~\bibnamefont {Weng}}, \bibinfo
  {author} {\bibfnamefont {Q.}~\bibnamefont {Wu}}, \bibinfo {author}
  {\bibfnamefont {X.}~\bibnamefont {Dai}},\ and\ \bibinfo {author}
  {\bibfnamefont {Z.}~\bibnamefont {Fang}},\ }\bibfield  {title} {\bibinfo
  {title} {Three-dimensional dirac semimetal and quantum transport in
  cd${}_{3}$as${}_{2}$},\ }\href {https://doi.org/10.1103/PhysRevB.88.125427}
  {\bibfield  {journal} {\bibinfo  {journal} {Phys. Rev. B}\ }\textbf {\bibinfo
  {volume} {88}},\ \bibinfo {pages} {125427} (\bibinfo {year}
  {2013})}\BibitemShut {NoStop}%
\bibitem [{\citenamefont {Borisenko}\ \emph {et~al.}(2014)\citenamefont
  {Borisenko}, \citenamefont {Gibson}, \citenamefont {Evtushinsky},
  \citenamefont {Zabolotnyy}, \citenamefont {B\"uchner},\ and\ \citenamefont
  {Cava}}]{PhysRevLett113027603}%
  \BibitemOpen
  \bibfield  {author} {\bibinfo {author} {\bibfnamefont {S.}~\bibnamefont
  {Borisenko}}, \bibinfo {author} {\bibfnamefont {Q.}~\bibnamefont {Gibson}},
  \bibinfo {author} {\bibfnamefont {D.}~\bibnamefont {Evtushinsky}}, \bibinfo
  {author} {\bibfnamefont {V.}~\bibnamefont {Zabolotnyy}}, \bibinfo {author}
  {\bibfnamefont {B.}~\bibnamefont {B\"uchner}},\ and\ \bibinfo {author}
  {\bibfnamefont {R.~J.}\ \bibnamefont {Cava}},\ }\bibfield  {title} {\bibinfo
  {title} {Experimental realization of a three-dimensional dirac semimetal},\
  }\href {https://doi.org/10.1103/PhysRevLett.113.027603} {\bibfield  {journal}
  {\bibinfo  {journal} {Phys. Rev. Lett.}\ }\textbf {\bibinfo {volume} {113}},\
  \bibinfo {pages} {027603} (\bibinfo {year} {2014})}\BibitemShut {NoStop}%
\bibitem [{\citenamefont {Sandvik}(2010)}]{10.1063/1.3518900}%
  \BibitemOpen
  \bibfield  {author} {\bibinfo {author} {\bibfnamefont {A.~W.}\ \bibnamefont
  {Sandvik}},\ }\bibfield  {title} {\bibinfo {title} {{Computational Studies of
  Quantum Spin Systems}},\ }\href {https://doi.org/10.1063/1.3518900}
  {\bibfield  {journal} {\bibinfo  {journal} {AIP Conference Proceedings}\
  }\textbf {\bibinfo {volume} {1297}},\ \bibinfo {pages} {135} (\bibinfo {year}
  {2010})}\BibitemShut {NoStop}%
\bibitem [{\citenamefont {Sandvik}(2007)}]{PhysRevLett.98.227202}%
  \BibitemOpen
  \bibfield  {author} {\bibinfo {author} {\bibfnamefont {A.~W.}\ \bibnamefont
  {Sandvik}},\ }\bibfield  {title} {\bibinfo {title} {Evidence for deconfined
  quantum criticality in a two-dimensional heisenberg model with four-spin
  interactions},\ }\href {https://doi.org/10.1103/PhysRevLett.98.227202}
  {\bibfield  {journal} {\bibinfo  {journal} {Phys. Rev. Lett.}\ }\textbf
  {\bibinfo {volume} {98}},\ \bibinfo {pages} {227202} (\bibinfo {year}
  {2007})}\BibitemShut {NoStop}%
\bibitem [{\citenamefont {Da~Liao}\ \emph {et~al.}(2019)\citenamefont
  {Da~Liao}, \citenamefont {Meng},\ and\ \citenamefont
  {Xu}}]{PhysRevLett.123.157601}%
  \BibitemOpen
  \bibfield  {author} {\bibinfo {author} {\bibfnamefont {Y.}~\bibnamefont
  {Da~Liao}}, \bibinfo {author} {\bibfnamefont {Z.~Y.}\ \bibnamefont {Meng}},\
  and\ \bibinfo {author} {\bibfnamefont {X.~Y.}\ \bibnamefont {Xu}},\
  }\bibfield  {title} {\bibinfo {title} {Valence bond orders at charge
  neutrality in a possible two-orbital extended hubbard model for twisted
  bilayer graphene},\ }\href {https://doi.org/10.1103/PhysRevLett.123.157601}
  {\bibfield  {journal} {\bibinfo  {journal} {Phys. Rev. Lett.}\ }\textbf
  {\bibinfo {volume} {123}},\ \bibinfo {pages} {157601} (\bibinfo {year}
  {2019})}\BibitemShut {NoStop}%
\bibitem [{\citenamefont {Da~Liao}\ \emph {et~al.}(2022)\citenamefont
  {Da~Liao}, \citenamefont {Xu}, \citenamefont {Meng},\ and\ \citenamefont
  {Qi}}]{PhysRevB.106.075111}%
  \BibitemOpen
  \bibfield  {author} {\bibinfo {author} {\bibfnamefont {Y.}~\bibnamefont
  {Da~Liao}}, \bibinfo {author} {\bibfnamefont {X.~Y.}\ \bibnamefont {Xu}},
  \bibinfo {author} {\bibfnamefont {Z.~Y.}\ \bibnamefont {Meng}},\ and\
  \bibinfo {author} {\bibfnamefont {Y.}~\bibnamefont {Qi}},\ }\bibfield
  {title} {\bibinfo {title} {{Dirac fermions with plaquette interactions. I.
  SU(2) phase diagram with Gross-Neveu and deconfined quantum criticalities}},\
  }\href {https://doi.org/10.1103/PhysRevB.106.075111} {\bibfield  {journal}
  {\bibinfo  {journal} {Phys. Rev. B}\ }\textbf {\bibinfo {volume} {106}},\
  \bibinfo {pages} {075111} (\bibinfo {year} {2022})}\BibitemShut {NoStop}%
\bibitem [{\citenamefont {Zerf}\ \emph {et~al.}(2017)\citenamefont {Zerf},
  \citenamefont {Mihaila}, \citenamefont {Marquard}, \citenamefont {Herbut},\
  and\ \citenamefont {Scherer}}]{PhysRevD.96.096010}%
  \BibitemOpen
  \bibfield  {author} {\bibinfo {author} {\bibfnamefont {N.}~\bibnamefont
  {Zerf}}, \bibinfo {author} {\bibfnamefont {L.~N.}\ \bibnamefont {Mihaila}},
  \bibinfo {author} {\bibfnamefont {P.}~\bibnamefont {Marquard}}, \bibinfo
  {author} {\bibfnamefont {I.~F.}\ \bibnamefont {Herbut}},\ and\ \bibinfo
  {author} {\bibfnamefont {M.~M.}\ \bibnamefont {Scherer}},\ }\bibfield
  {title} {\bibinfo {title} {Four-loop critical exponents for the
  gross-neveu-yukawa models},\ }\href
  {https://doi.org/10.1103/PhysRevD.96.096010} {\bibfield  {journal} {\bibinfo
  {journal} {Phys. Rev. D}\ }\textbf {\bibinfo {volume} {96}},\ \bibinfo
  {pages} {096010} (\bibinfo {year} {2017})}\BibitemShut {NoStop}%
\bibitem [{\citenamefont {Moessner}\ \emph {et~al.}(2001)\citenamefont
  {Moessner}, \citenamefont {Sondhi},\ and\ \citenamefont
  {Chandra}}]{PhysRevB.64.144416}%
  \BibitemOpen
  \bibfield  {author} {\bibinfo {author} {\bibfnamefont {R.}~\bibnamefont
  {Moessner}}, \bibinfo {author} {\bibfnamefont {S.~L.}\ \bibnamefont
  {Sondhi}},\ and\ \bibinfo {author} {\bibfnamefont {P.}~\bibnamefont
  {Chandra}},\ }\bibfield  {title} {\bibinfo {title} {Phase diagram of the
  hexagonal lattice quantum dimer model},\ }\href
  {https://doi.org/10.1103/PhysRevB.64.144416} {\bibfield  {journal} {\bibinfo
  {journal} {Phys. Rev. B}\ }\textbf {\bibinfo {volume} {64}},\ \bibinfo
  {pages} {144416} (\bibinfo {year} {2001})}\BibitemShut {NoStop}%
\bibitem [{\citenamefont {Harada}(2011)}]{PhysRevE.84.056704}%
  \BibitemOpen
  \bibfield  {author} {\bibinfo {author} {\bibfnamefont {K.}~\bibnamefont
  {Harada}},\ }\bibfield  {title} {\bibinfo {title} {Bayesian inference in the
  scaling analysis of critical phenomena},\ }\href
  {https://doi.org/10.1103/PhysRevE.84.056704} {\bibfield  {journal} {\bibinfo
  {journal} {Phys. Rev. E}\ }\textbf {\bibinfo {volume} {84}},\ \bibinfo
  {pages} {056704} (\bibinfo {year} {2011})}\BibitemShut {NoStop}%
\bibitem [{\citenamefont {Hertz}(1976)}]{PhysRevB.14.1165}%
  \BibitemOpen
  \bibfield  {author} {\bibinfo {author} {\bibfnamefont {J.~A.}\ \bibnamefont
  {Hertz}},\ }\bibfield  {title} {\bibinfo {title} {Quantum critical
  phenomena},\ }\href {https://doi.org/10.1103/PhysRevB.14.1165} {\bibfield
  {journal} {\bibinfo  {journal} {Phys. Rev. B}\ }\textbf {\bibinfo {volume}
  {14}},\ \bibinfo {pages} {1165} (\bibinfo {year} {1976})}\BibitemShut
  {NoStop}%
\bibitem [{\citenamefont {Millis}(1993)}]{PhysRevB.48.7183}%
  \BibitemOpen
  \bibfield  {author} {\bibinfo {author} {\bibfnamefont {A.~J.}\ \bibnamefont
  {Millis}},\ }\bibfield  {title} {\bibinfo {title} {Effect of a nonzero
  temperature on quantum critical points in itinerant fermion systems},\ }\href
  {https://doi.org/10.1103/PhysRevB.48.7183} {\bibfield  {journal} {\bibinfo
  {journal} {Phys. Rev. B}\ }\textbf {\bibinfo {volume} {48}},\ \bibinfo
  {pages} {7183} (\bibinfo {year} {1993})}\BibitemShut {NoStop}%
\bibitem [{\citenamefont {Vojta}(2003)}]{MatthiasVojta_2003}%
  \BibitemOpen
  \bibfield  {author} {\bibinfo {author} {\bibfnamefont {M.}~\bibnamefont
  {Vojta}},\ }\bibfield  {title} {\bibinfo {title} {Quantum phase
  transitions},\ }\href {https://doi.org/10.1088/0034-4885/66/12/R01}
  {\bibfield  {journal} {\bibinfo  {journal} {Reports on Progress in Physics}\
  }\textbf {\bibinfo {volume} {66}},\ \bibinfo {pages} {2069} (\bibinfo {year}
  {2003})}\BibitemShut {NoStop}%
\bibitem [{\citenamefont {Sachdev}(1999)}]{Sachdev_1999}%
  \BibitemOpen
  \bibfield  {author} {\bibinfo {author} {\bibfnamefont {S.}~\bibnamefont
  {Sachdev}},\ }\bibfield  {title} {\bibinfo {title} {Quantum phase
  transitions},\ }\href {https://doi.org/10.1088/2058-7058/12/4/23} {\bibfield
  {journal} {\bibinfo  {journal} {Physics World}\ }\textbf {\bibinfo {volume}
  {12}},\ \bibinfo {pages} {33} (\bibinfo {year} {1999})}\BibitemShut {NoStop}%
\bibitem [{\citenamefont {XU}(2012)}]{S0217979212300071}%
  \BibitemOpen
  \bibfield  {author} {\bibinfo {author} {\bibfnamefont {C.}~\bibnamefont
  {XU}},\ }\bibfield  {title} {\bibinfo {title} {Unconventional quantum
  critical points},\ }\href {https://doi.org/10.1142/S0217979212300071}
  {\bibfield  {journal} {\bibinfo  {journal} {International Journal of Modern
  Physics B}\ }\textbf {\bibinfo {volume} {26}},\ \bibinfo {pages} {1230007}
  (\bibinfo {year} {2012})}\BibitemShut {NoStop}%
\end{thebibliography}%

\end{document}